\documentclass{aa}  
\usepackage{graphicx}
\usepackage{scalefnt}
\usepackage{txfonts,textcomp}
\usepackage{hyperref}
\newbox\grsign \setbox\grsign=\hbox{$>$} \newdimen\grdimen \grdimen=\ht\grsign
\newbox\simlessbox \newbox\simgreatbox
\setbox\simgreatbox=\hbox{\raise.5ex\hbox{$>$}\llap
     {\lower.5ex\hbox{$\sim$}}}\ht1=\grdimen\dp1=0pt
\setbox\simlessbox=\hbox{\raise.5ex\hbox{$<$}\llap
     {\lower.5ex\hbox{$\sim$}}}\ht2=\grdimen\dp2=0pt

\def\simless{\mathrel{\copy\simlessbox}}
\newbox\simppropto
\setbox\simppropto=\hbox{\raise.5ex\hbox{$\sim$}\llap
     {\lower.5ex\hbox{$\propto$}}}\ht2=\grdimen\dp2=0pt

\begin{document}

\title{\fontsize{14.3}{10}\selectfont{Abundances of  P, S, and K in 58 bulge spheroid stars from APOGEE}}
\titlerunning{Abundances of  P, S, and K in 58 bulge spheroid stars from APOGEE}
 \authorrunning{B. Barbuy et al.}
 
 \author{\fontsize{10}{10}\selectfont{
 B. Barbuy \inst{1} 
\and H. Ernandes \inst{2}
 \and A.C.S. Fria\c ca\inst{1} 
  \and M.S. Camargo\inst{1} 
  \and P. da Silva \inst{1} 
 \and S. O. Souza \inst{3} 
 \and T. Masseron\inst{4,5}
  \and M. Brauner\inst{4,5}
  \and D. A. Garc\'{\i}a-Hern\'andez\inst{4,5}
 \and  J. G. Fern\'andez-Trincado \inst{6}  
\and K. Cunha\inst{7,8} 
\and V. V. Smith\inst{9}
\and A. P\'erez-Villegas\inst{10}
\and C. Chiappini\inst{11}
\and A.B.A. Queiroz\inst{4}
\and B. X. Santiago\inst{12}
\and T. C. Beers\inst{13}
\and F. Anders\inst{14,15,16}
\and R. P. Schiavon\inst{17}
\and M. Valentini\inst{11}
\and D. Minniti\inst{18,19}
\and D. Geisler\inst{20,21}
\and D. Souto\inst{22}
\and V. M. Placco\inst{9}
\and M. Zoccali\inst{23}
\and S. Feltzing\inst{2}
\and M. Schultheis\inst{24}
\and C. Nitschelm\inst{25}
}
}

\institute{
Universidade de S\~ao Paulo,  IAG, Departamento de Astronomia, 05508-090 S\~ao Paulo, Brazil
\and 
Lund Observatory, Department of Astronomy and Theoretical Physics, Lund University, Box 43, SE-221 00 Lund, Sweden 
\and 
Max Planck Institute for Astronomy, K\"onigstuhl 17, D-69117 Heidelberg, Germany
\and
 Instituto de Astrof\'{\i}sica de Canarias, C/Via Lactea s/n, E-38205 La Laguna, Tenerife, Spain 
\and
Departamento de Astrof\'{\i}sica, Universidad de La Laguna, E-38206 La Laguna, Tenerife, Spain
\and
Instituto de Astronom\'ia, Universidad Cat\'olica del Norte, Av. Angamos 0610, Antofagasta, Chile
\and
University of Arizona, Steward Observatory, Tucson, AZ 85719, USA 
 \and
Observat\'orio Nacional,  rua General Jos\'e  Cristino 77, S\~ao Crist\'ov\~ao, Rio de Janeiro 20921-400, Brazil
\and
NSF NOIRLab, 950 N. Cherry Ave., Tucson, AZ 85719, USA 
\and
Instituto de Astronom\'ia, Universidad Nacional Aut\'onoma de M\'exico, A. P. 106, C.P. 22800, Ensenada, B. C., M\'exico
\and
Astrophysikalisches Institut Potsdam, An der Sternwarte 16, Potsdam, 14482, Germany
\and
Universidade Federal do Rio Grande do Sul, Caixa Postal 15051, 91501-970 Porto Alegre, Brazil
\and
Department of Physics and Astronomy and JINA Center for the Evolution of the Elements (JINA-CEE), University of Notre Dame, Notre Dame, IN 46556  USA
\and
Departament de F\'{\i}sica Qu\~antica i Astrof\'{\i}sica (FQA), Universitat de Barcelona (UB), Mart\'{\i} i Franqu\`es, 1, 08028 Barcelona, Spain
\and
 Institut de Ci\`encies del Cosmos, Universitat de Barcelona (IEEC-UB), Mart\'{\i} i Franqu\`es 1, 08028 Barcelona, Spain
\and
Institut d'Estudis Espacials de Catalunya (IEEC), Edifici RDIT, Campus UPC, 08860 Castelldefels (Barcelona), Spain
 \and
Astrophysics Research Institute, Liverpool John Moores University, Liverpool, L3 5RF, UK 
\and 
Instituto de Astrof\'isica, Facultad de Ciencias Exactas, Universidad Andres Bello, Fern\'andez Concha 700, Las Condes, Santiago, Chile 
\and 
Vatican Observatory, Vatican City State 00120, Italy 
\and
Departamento de Astronomia, Casilla 160-C, Universidad de Concepcion, Chile 
\and
Departamento de Astronomía, Facultad de Ciencias, Universidad de La Serena. Av.
Avenida Ra\'ul Bitr\'an S/N, La Serena, Chile 
 \and
 Universidade Federal de Sergipe, Av. Marechal Rondon, S/N, 49000-000
S\~ao Crist\'ov\~ao, SE, Brazil
\and
Instituto de Astrof\'isica, Pontificia Universidad Cat\'olica de Chile, Vicu\~na Mackenna 4860, Macul, Casilla 306, Santiago 22, Chile
 \and
 Universit\'e C\^ote d'Azur, Observatoire de la C\^ote d'Azur, CNRS, Laboratoire Lagrange, Nice, France 
 \and
Centro de Astronom{\'i}a (CITEVA), Universidad de Antofagasta, Avenida Angamos 601, Antofagasta 1270300, Chile 
}
            
   \date{Received ....; accepted .....}
 
  \abstract
{ We have previously studied several elements in 58 selected bulge spheroid stars, based on spectral lines in the \textit{H}-band. We now derive the abundances  of  the less-studied elements phosphorus
 (P; Z=15), sulphur (S; Z=16), and potassium (K; Z=19).}
   { The abundances of P, S, and K in 58 bulge spheroid stars are compared both with the results of a  previous analysis of the data from the Apache Point Observatory Galactic Evolution Experiment (APOGEE), and with a few available studies of these elements.}
  {We derive the individual abundances through spectral synthesis, using the stellar physical parameters available for our sample from the DR17 release of the APOGEE project. We provide recommendations for the best lines to be used for the studied elements among those in the {\it H}-band. We also compare the present results, together with literature data, with chemical-evolution models. Finally, the neutrino-process was taken into account for the suitable fit to the odd-Z elements P and K.} 
   {We confirm that the \textit{H}-band has useful lines for the
   derivation of the elements P, S, and K in moderately
   metal-poor stars. The abundances, plotted together with literature results from high-resolution spectroscopy, indicate that: moderately enhanced phosphorus stars are found, reminiscent results obtained for thick disk and halo stars of metallicity [Fe/H]$\approx$$-$1.0. Therefore, for the first time, we identify this effect to occur in the old stars from the bulge spheroid.
   Sulphur is an $\alpha$-element and behaves as such.
   Potassium and sulphur both exhibit some star-to-star scatter, but fit within the expectations from chemical evolution models.
   
   }
  {}
   \maketitle

\section{Introduction}

Galaxy bulges and inner haloes form first \citep[e.g.][]{gao10}, and
the oldest stars in the Milky Way were very likely formed in the early bulge. 
We have identified a sample of 58 stars that appear to represent the stellar population
of an early bulge spheroid, based on chemical, kinematical, and dynamical criteria,
as described in \citet{razera22}.
These old bulge stars were selected to have metallicities  of [Fe/H]$<-$0.8, 
chosen to exclude most of bulge stars that are metal-rich, and to include stars from a small metallicity peak at [Fe/H]$\sim-$1.0,
that is detected among both bulge globular clusters \citep{rossi15,bica16,perez-villegas20,bica24}
and field stars \citep{lucey21}. This relatively high metallicity for the oldest stars
is due to rapid chemical enrichment in the central region of the Galaxy \citep{Chiappini11,wise12,friaca17,barbuy18a,matteucci21}. 
There is some evidence that many of the stars located in the bulge with metallicity
[Fe/H]$\simless$$-$1.5 should be assigned to a halo origin, as discussed in \citet{lucey21},
and also suggested by \citet{fernandez-trincado20} and \citet{geisler23} as concerns the bulge globular clusters.

The kinematical characteristics of the sample stars 
together with analysed abundances of C, N, O, Mg, Si, Ca, and Ce are reported in \citet{razera22}. \citet{barbuy23}, 
and  \citet{barbuy24} derived Na and Al, and iron-peak elements  V, Cr, Mn, Co, Ni, and Cu, respectively, and interpreted these abundances
in terms of their chemical-evolution models.  In \citet{sales-silva24} the neutron-capture elements Nd and Ce,
including some stars in common with the present sample, were analysed.
In this paper, we analyze the less studied elements P, S, and K. These studies are based on data from the
Apache Point Observatory Galactic Evolution Experiment \citep[APOGEE;][]{majewski17}.

The elements P, S, and K are produced in supernovae type II (SNII), also called core-collapse supernovae
(CCSNe) \citep[e.g.][hereafter WW95]{woosley95}. 
\citet{cescutti12} chemical evolution models succeeded in reproducing P in disk stars data by using the massive star yields of \citet{kobayashi06} multiplied by a factor of 2.75. 
 Based on data for halo stars, \citet{jacobson14} favored the yields from
 hypernovae by \citet{kobayashi06} to reproduce their data.
These elements were little studied in the literature, primarily because they possess only a
few measurable and unblended spectral lines in the optical. For phosphorus,
the {\it H}-band shows one measurable and a second fainter line. Sulphur has lines in the region 600-900nm, that are non-negligibly affected by
NLTE, and one measurable line in the {\it H}-band. Potassium has two lines in the {\it H}-band, both only blended with rather weak CN festures. In the present work we
derive the abundances of these elements using spectrum synthesis.

This paper is organized as follows. In Section \ref{sample}, we summarize the selection of sample stars and the available data.
In Section \ref{calculations}, we describe the calculations for spectral synthesis and the lines studied. In Section \ref{sec:results}, we present the results, along with the data from the literature on the elements studied.
In Section \ref{models} the chemical-evolution models are compared with the data. 
We summarise our conclusions in Section \ref{conclusions}.

\section{Data and selection sample}\label{sample}

As described in \citet{razera22}, our selection is based on the
reduced proper motion (RPM) stars from \citet{queiroz21} that served as a pool to select the sample. 
The RPM sample from \cite{queiroz21}, in turn,  is based on the stars observed by APOGEE, combined with {\tt{StarHorse}} distances \citep{santiago16, queiroz18}, and cross-matched with proper motions from the  Gaia Early Data Release 3 \citep{gaia21}. The selection identified stars with a distance to the Galactic Center of d$_{\rm GC}< 4$ kpc,  a maximum height of $| Z|_{\rm max}< 3$ kpc, eccentricity $> 0.7$, and with orbits not supporting the bar (where the orbits were computed in \citealt{queiroz21}),
and imposing a metallicity cut of [Fe/H] $<-0.8$.
By applying kinematical criteria and imposing that the selected stars have spectra from the second generation of the 
APOGEE-2 - \citealt{majewski17}, a sample of 58 stars was build. A Kiel diagram of this sample was presented in Fig. 3 of \citet{razera22}.

APOGEE is one of the surveys of the Sloan Digital Sky Survey IV \citep[SDSS-IV/V]{blanton17}. The APOGEE
spectra have a high resolution ($R \sim$ 22,500)  and
high signal-to-noise ratios in the \textit{H}-band  (15140-16940 {\rm \AA}) \citep{wilson19}, and include 
about 7$\times$10$^{5}$ stars.
 APOGEE-1  and APOGEE-2 used the 2.5m Sloan Foundation Telescope at the Apache Point Observatory in New Mexico 
 \citep{gunn06}, and the 2.5m Ir\'en\'ee du Pont Telescope at the Las Campanas Observatory in Chile \citep{bowen73}, respectively. 
 \citet{santana21} and \citet{beaton21} describe the targeting of  APOGEE
for the Southern and Northern Hemispheres, respectively. 
The detectors are H2RG (2048 x 2048) Near-Infrared HgCdTe Detectors with 18 micron pixels.

The analysis of \textit{H}-band spectra in the APOGEE project is carried out through a Nelder-Mead algorithm 
\citep{nelder65},
which simultaneously  fits the stellar parameters effective temperature (T$_{\rm eff}$), gravity (log~g), 
metallicity ([Fe/H]), and microturbulence velocity (v$_{\rm t}$) -- together with the abundances of carbon, nitrogen, and $\alpha$-elements with the APOGEE Stellar Parameter and Chemical
Abundances Pipeline (ASPCAP) \citep{garcia-perez16}, which is based on the FERRE code \citep{allende-prieto06} and the APOGEE line list \citet{smith21}. 
In the present work, we use APOGEE Data Release 17 - DR17 \citep{abdurro22} for the stellar parameters,
to derive the abundances with spectrum synthesis, in particular for phosphorus, that are not available from DR17;
The P abundance is available for uniquely one star in APOGEE DR17, to be discussed in Sect. \ref{uncertainties}.

\section{Calculations}\label{calculations}

We computed the abundances of P, S, and K in the \textit{H}-band
 using the code {\tt TURBOSPECTRUM} from \citet{alvarez98} and \citet{plez12}. 
The model atmosphere are interpolated within the CN-mild MARCS grids from \citet{gustafsson08}. 
The solar abundances of the elements studied are from \citet{asplund21}, that is, $A$(P) = 5.41, $A$(S) = 7.12 and $A$(K) = 5.07. Note that, from O and B stars,  recently
\citet{aschenbrenner25} derived a present-day abundance for P,  of A(P) = 
5.36$\pm$0.14, and there are present-day abundances for S, derived by \citet{daflon09}, of A(S) = 7.15$\pm$0.05.
We also recomputed the C, N, O abundances, given that in \citet{razera22} these
abundances were computed with a different code PFANT \citet{barbuy18b}. 
Here we assumed the solar C, N, O abundances of \citet{asplund21}, A (C) = 8.46, A (N) = 7.83,
and A(O) = 8.69, also different solar values from those assumed in \citet{razera22}.

Table \ref{linelist} reports the lines in the \textit{H}-band that we used to measure the abundances
of the elements P, S, and K in the spectra of the sample stars. 
Oscillator strengths were adopted from the line list of the APOGEE collaboration,
which were initially adopted from  the most recent line
list of \citet{kurucz95}, 
with log gf values updated with National Institute of Standards and Technology Atomic Spectra Database - NIST values, and
adjusted within 2-sigma uncertainties based on the Sun and Arcturus spectra
\citep{smith21}.
For comparison purposes, we also show the log~gf values 
from the Kurucz CD ROM 23, and from the Vienna Atomic Line Database (VALD3): see the  line lists of 
\citet{kurucz95} and \citet{ryabchikova15}. We note that log~gf values are not available in the NIST database
for any of these studied lines. 
The  full atomic line list employed is that from the APOGEE collaboration,
together with the molecular lines described in \citet{smith21}. 

\begin{table}
\small
\scalefont{0.5}
\centering
\caption[4]{Line list and oscillator strengths. }
\resizebox{0.4\textwidth}{!}{
\begin{tabular}{l@{}ccrr@{}rcccccccc}
\hline
\hline
\noalign{\smallskip}
\hbox{Ion} & \hbox{$\lambda$} & \hbox{$\chi_{ex}$}  &\hbox{log~gf}  &\hbox{log~gf}  &\hbox{log~gf} \\
& \hbox{(\AA)} &\hbox{(eV)} & \hbox{VALD3}   & \hbox{Kurucz} & \hbox{APOGEE}    \\ 
\noalign{\smallskip}
\hline
\noalign{\smallskip}
\hbox{PI} 
& 15711.522 & 7.176 & $-$0.510 & $-$0.720 & $-$0.404\\
& 16482.932 & 7.213 & $-$0.290 & $-$0.400 & $-$0.273 \\ 
\hbox{SI} & 15469.816 & 8.046 &$-$0.050 & $-$0.220 & $-$0.199  \\
& 15475.616 & 8.047 & $-$0.520 & $-$0.700  & $-$0.744 \\ 
& 15478.482 & 8.047 & 0.180 & 0.000 & $-$0.040 \\ 
\hbox{KI} & 15163.067 & 2.670 & 0.524 & 0.640 & 0.630 \\
& 15168.377 & 2.670 & 0.347 & 0.480 & 0.481 \\
\hline
\noalign{\hrule\vskip 0.1cm} 
\hline                 
\label{linelist}
\end{tabular}}
\begin{minipage}{8cm}
\vspace{0.1cm}
\small Note: Air wavelengths are from APOGEE collaboration linelist. Note that somewhat different wavelengths
for \ion{S}{I} 15469.826. 15475.624 and 15478.496 {\rm \AA} are given in National Institute of Standards and Technology (NIST) by
Kramida, A., Ralchenko, Yu., Reader, J., and the NIST ASD Team (2024). The NIST Atomic Spectra Database used is (ver. 5.12), [Online]\footnote{https://physics.nist.gov/asd} [2025, April 16].

Oscillator strengths of the P, S, and K lines from VALD3,  \citet{kurucz95}, 
and the APOGEE collaboration (adopted) are reported.
\end{minipage}
\end{table} 

We adopted the uncalibrated stellar parameters 
effective temperature (T$_{\rm eff}$), gravity (log~g), metallicity ([Fe/H]), and microturbulence velocity (v$_{t}$) from the APOGEE DR17 results. 
These parameters are reported in Table \ref{results}, together with 
recomputed C, N, O,  and the results on P, S, and K abundances.

The use of uncalibrated instead of the calibrated parameters could be a matter of debate because the difference in parameters can lead to changes in the abundances derived. While using calibrated parameters may seem to be a more sensible choice for most of the giants of the APOGEE survey, we argue here that this may not be as obvious for our current sample. According to \citet{Jonsson2020}, ASPCAP calibrated temperatures have been calibrated using the color-temperature relation of \citet{Gonzalez2009} which has been tested down to 4000\,K, and surface gravities have been calibrated using seismic data from \citet{Pinsonneault2018} and validated down to $\log g =$1.2. Because our sample, such as reported here in Table \ref{results}, contains only very cool (3600K-4300K) metal-poor ([Fe/H]$\sim$-1.0) low gravity ($\log g \lesssim$1.2) giants, it is not clear anymore whether the calibrated parameters perform better than the uncalibrated parameters. In addition, we also use a grid of model atmosphere slightly different in C, N and alpha-element composition than APOGEE-ASPCAP. For the cooler stars (T$_{\rm eff}$) $\lesssim$ 3700\,K), that will also have a significant impact in the atmosphere structure, thus on parameters and abundances. However, neither the MARCS atmosphere grid we use here, nor the APOGEE-ASPCAP one strictly matches the abundances of our sample stars and it is not clear which of them is best for such sample. We still chose the uncalibrated parameters in order to maintain some consistency with the raw results of the APOGEE-ASPCAP pipeline, as previously discussed in \citet{razera22}.
We also note that in  \citet{dasilva24} we verified the reliability of ASPCAP for deriving stellar parameters, and concluded that the use of molecular-line
intensities is a powerful method, in particular for the derivation
of effective temperatures.
We remind that for DR17, the parameters were obtained with new spectral grids constructed using the Synspec spectrum synthesis code \citep{hubeny17,hubeny21}.

\subsection{Phosphorus, sulphur, and potassium lines}\label{lines}

We analyzed  the lines of P, S, and K in the {\it H}-band. The fits
were all carried out visually, adopting convolutions with 
full width at half maximum (FWHM) from 0.65 to 0.75 {\rm \AA} in 
the range 15,000 to 17,000 {\rm \AA}. These FWHM values are compatible with
those based on a directly measured FWHM of $\sim$0.7 {\rm \AA}, with 10 to 20 per cent
variations seen across the wavelength range by \citet{ashok21} and \citet{nidever15}.
Although for most of the lines the resulting abundance is similar to that reported
in APOGEE-ASPCAP DR17, there are cases where the visual inspection is needed because of noise or defects. 
The results are given in Table \ref{results}.

{\it Phosphorus}: The two available lines of \ion{P}{I} are weak. The 
\ion{P}{I} 15711.522 {\rm \AA} is often too weak, while \ion{P}{I} 16482.932 {\rm \AA}
is measurable. In about one fourth of the stars, the \ion{P}{I} 15711.522 {\rm \AA}  line is measurable, and with an abundance compatible with that of the \ion{P}{I} 16482.932 {\rm \AA} line. The upper limit of the 15711.522 {\rm \AA}  line is compatible with the measurement of the 16482.932 {\rm \AA} line for the other two third of stars. We also stress that P measurement was not possible in the spectra where the 16482.932 {\rm \AA} line was too close to the continuum, or too noisy and with strong artifacts - or 
even falling out of range given the detector edge. 

\begin{figure}
    \centering
    \includegraphics[width=8.5cm]{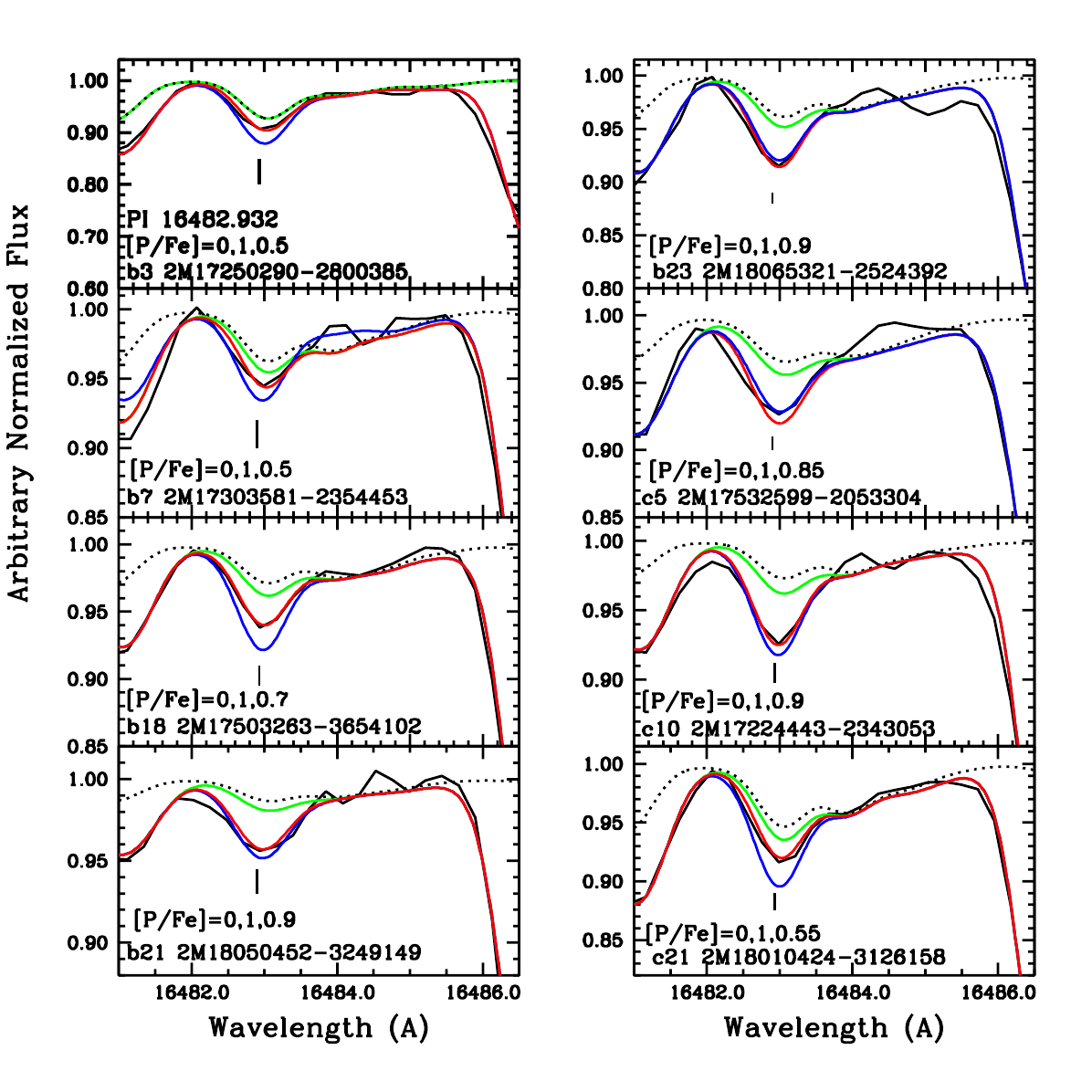}
  \caption{\ion{P}{I} 16482.932 {\rm \AA} line in 8 sample stars, fitted with synthetic spectra
   computed with  [P/Fe] = 0.0 (green), 1.0 (blue), and final values (red), if different from 0.0 or 1.0.
   The computation with molecular lines only are shown as dotted black lines.}
    \label{bestp}
\end{figure}

A very important issue is the mixing of the main \ion{P}{I} 16482.932 {\rm \AA} line with the molecular lines of CO, as noted by \citet{hayes22} and \citet{brauner23}.
Although we have already derived C, N, and O abundances for these stars in \citet{razera22}, it is crucial to adjust precisely the strength of the related molecular features to our current methodology. For that we basically use the region 15520 - 15590 {\rm \AA}, that contains a clear CO band-head, clean OH lines, and CN lines spread in this region,  as explained in \citet{barbuy21b} and \citet{razera22}.
Then, with CO-band lines well-fitted, the resulting estimate of
P becomes reliable.
Figure \ref{bestp} shows the \ion{P}{I} 16482.932 {\rm \AA} in the eight stars with the largest P abundances. The blending with molecular lines is also shown.

\begin{figure}
    \centering
    \includegraphics[width=8.5cm]{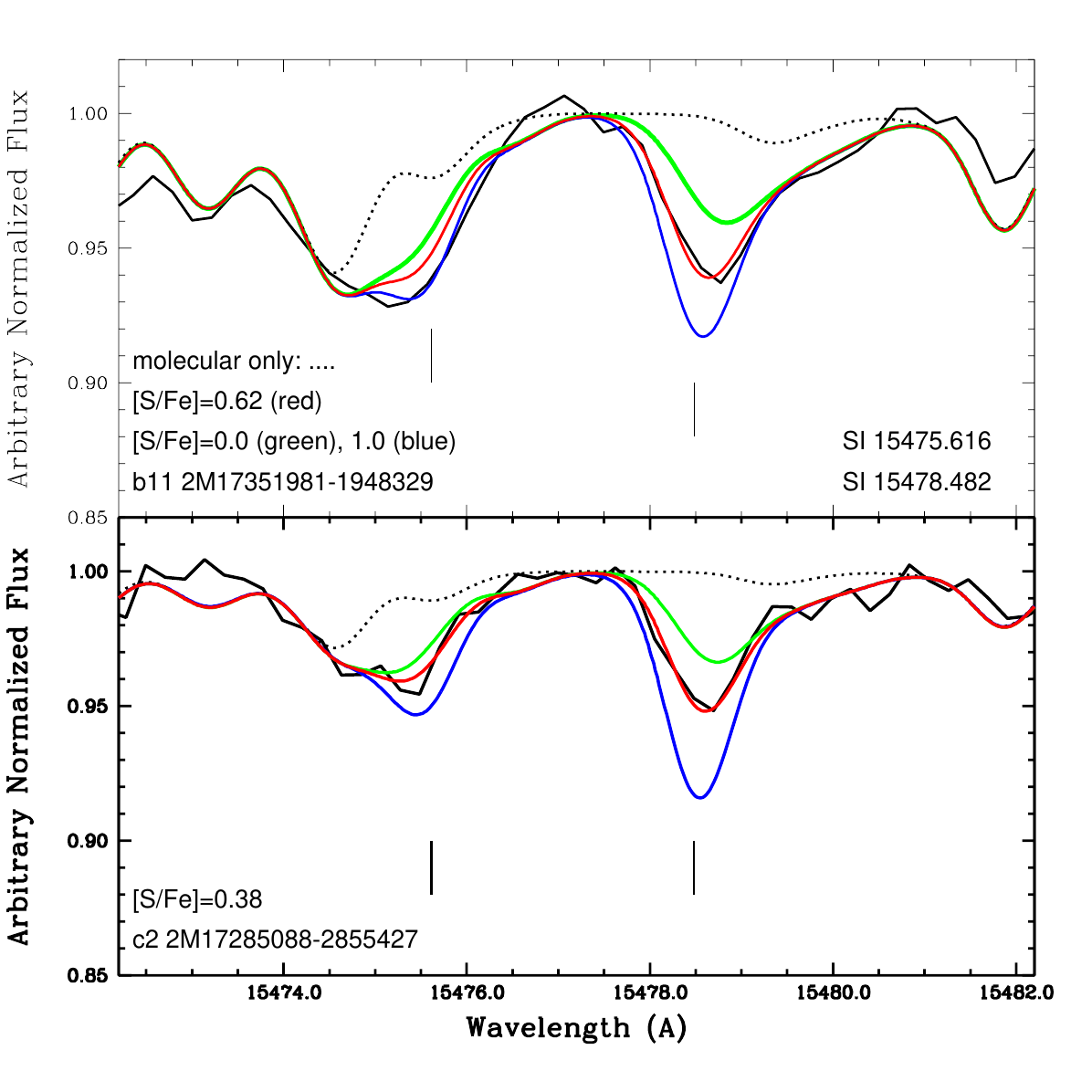}
  \caption{\ion{S}{I} 15475.624 and 15478.496 {\rm \AA} lines in stars
  b11 2M17351981-1948329 and c2 2M17285088-2855427. Observed spectra (black) are 
  compared with synthetic spectra
   computed with  [S/Fe] = 0.0 (green), 1.0 (blue), and final values (red).
   Dotted lines correspond to molecular lines only.}
    \label{figs}
\end{figure}

{\it Sulphur}: There are three measurable lines, but only one is unblended.
\ion{S}{I} 15403.784 {\rm \AA} is strongly blended with molecular OH lines, plus other atomic lines, 
and this applies also to 15469.826 {\rm \AA}, but the latter is still useful,
as discussed in \citet{hayes22}. 
\ion{S}{I} 15475.624  {\rm \AA} is  reasonably sensitive to the S abundance but is non-negligibly blended with CN.
Also, a few  other \ion{S}{I} lines fall in the APOGEE gaps.
The more reliable feature  that is  essentially clean from molecular and atomic blending, is the line \ion{S}{I}  15478.406 {\rm \AA}.
Fits to this feature are shown in Figure \ref{figs} for stars  b11 2M17351981-1948329 and c2 2M17285088-2855427, including the calculations with only molecular lines. Even if this line had more weight, the three features
were fitted simultaneously, i.e., 15469.826, 15475.624   and  15478.496 {\rm \AA}, and in part of the stars these features were reasonably fitted.

\begin{figure} 
    \centering
    \includegraphics[width=8.5cm]{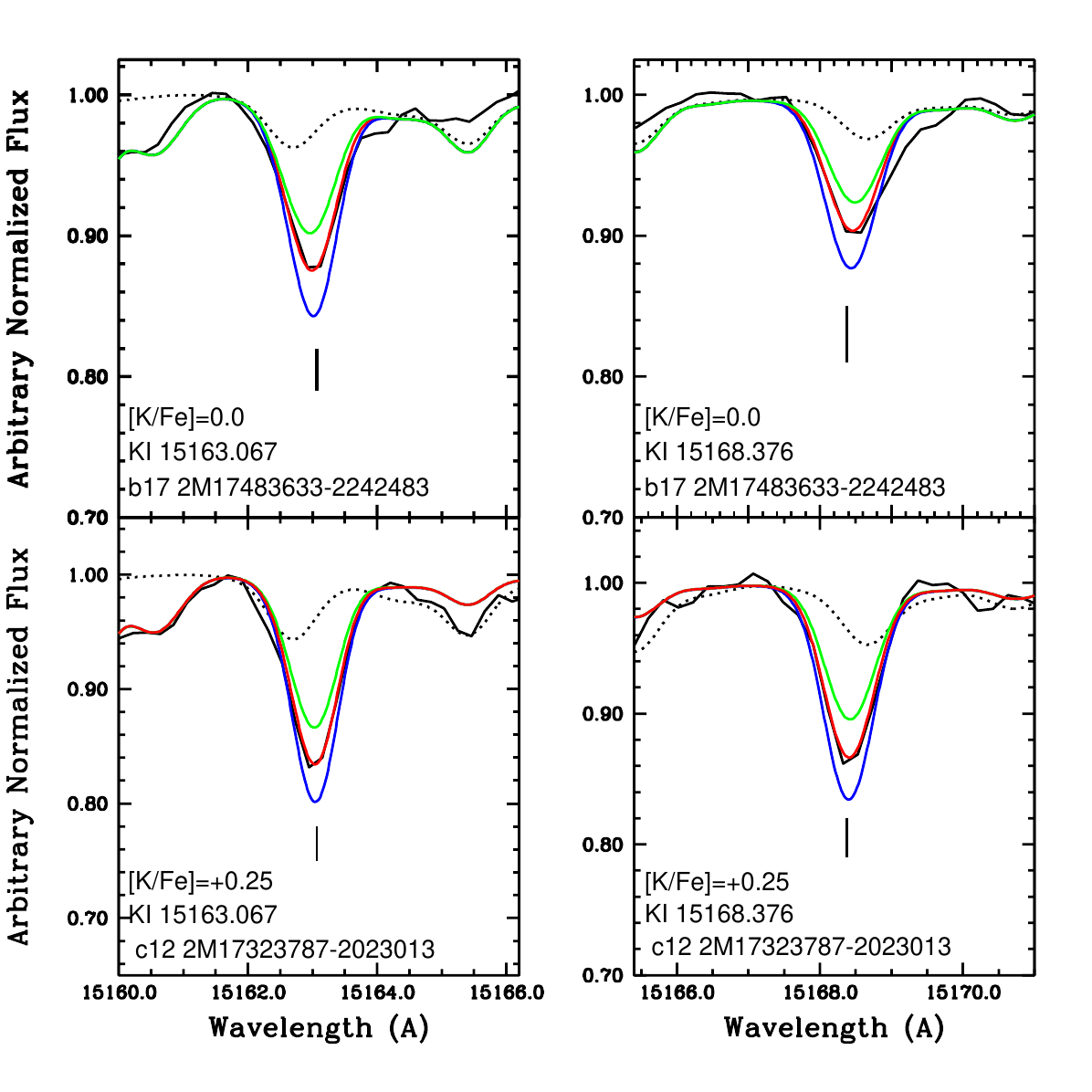}
  \caption{\ion{K}{I}  15163.067 and 15168.376 {\rm \AA} lines in the sample stars
  b17 2M17483633-2242483 and c12 2M17323787-2023013. The observed spectra (black)
  are fitted with synthetic spectra computed with 
   [K/Fe] = 0.0 for b17 and +0.25 for c12 respectively (red); calculations for -0.2 and +0.2 and shown in green and blue 
   respectively. Dotted lines correspond to molecular lines only.}
    \label{figk}
\end{figure}

{\it Potassium}: The two available lines are clearly measurable; both have a small blend with CN lines,
and they both yield results compatible with each other. Figure \ref{figk} illustrates the quality of these lines,
as well as the blending CN features, for stars
b17 2M17483633-2242483 and c12 2M17323787-2023013.

Table \ref{results} lists the APOGEE uncalibrated stellar parameters, and the present resulting abundances for the elements P, S, and K, as well as the APOGEE-ASPCAP abundances of S and K, for comparison purposes. 

\subsection{Uncertainties}\label{uncertainties}
There are no calculations of non-LTE deviations for the lines of 
\ion{P}{I}   analysed in this work, therefore these
uncertainties cannot be evaluated. 
\citet{brauner23} do not find any trend of P abundances versus temperature, suggesting that NLTE effects for the P lines in the $H$-band should be negligible.

For \ion{K}{I} the synspec
calculations from APOGEE DR17 take into account NLTE; since there
are available also the LTE calculations, we inspected the differences
between the two, and they are negligible, below 0.1 dex.

For sulphur, the large spread in the abundance estimates may
be improved with NLTE corrections, as it is for the case 
of lines studied by
\citet{korotin17}, lowering  the LTE abundances. \citet{korotin25} give grids of NLTE corrections
for the lines studied here, indicating corrections: as examples,
for [Fe/H] $\sim$ $-$1.0, T${\rm eff}$ = 4000 K, and 
[S/Fe] = +0.4 the correction is of $\sim$ 0.16 dex for 
a gravity log~g = 0.0, and $\sim$ $-$0.06 dex for log~g = 1.0; for [S/Fe] = +0.8, the corrections are of $\sim$ $-$0.19 and $-$0.07, respectively.
The grid is limited to T${\rm eff}$ $\geq$ 4000 K, and the trend for lower temperatures is a decrease of the correction, whereas
for higher effective temperature the corrections increase. In summary, for the low temperatures of most of our sample stars, the corrections should
be $\leq$ $-$0.06 dex.

Typical systematic uncertainties  due to stellar parameters are computed by adopting uncertainties in the stellar parameters of $\Delta$T$_{\rm eff}$ = 100\,K, $\Delta$log \textit{g} $=$ 0.2\,dex, 
$\Delta v_{\rm t}$ = 0.2 km $s^{-1}$, shown in Table \ref{errors}. This is applied to the cool star 
b14 2M17392719-2310311, and the warmer star c16 2M17310874-2956542.
This table shows that the  P abundance has a dependence on log~g, more so for the cooler star, and a low dependence on the effective temperature. Finally, the continuum placement is another source of uncertainty, in particular for the \ion{P}{I} line; this adds another
0.1\,dex of uncertainty for the noisier spectra.
Finally, it is worth nothing that this study of systematic errors offers a good representation of the impact on the abundances due to the use of either uncalibrated or calibrated parameters of APOGEE-ASPCAP DR17 as the difference between those parameters is on average similar to those used in Table \ref{errors}. 

\begin{table}
  \caption{Phosphorus, sulphur and potassium abundance uncertainties for stars b14 and c16, due to changes in stellar parameters of $\Delta$T$_{\rm eff}$ = 100 K, $\Delta$log \textit{g} $=$ 0.2\,dex, and $\Delta$v$_{\rm t}$ = 0.2 km s$^{-1}$. The corresponding total error is given in the last column. 
} 
\label{errors}
\begin{flushleft}
\small
\tabcolsep 0.5cm
\begin{tabular}{lc@{}c@{}c@{}c@{}c@{}}
\noalign{\smallskip}
\hline
\noalign{\smallskip}
\hline
\noalign{\smallskip}
\hbox{Element} & \hbox{$\Delta$T} & \hbox{$\Delta$log $g$} & 
\phantom{-}\hbox{$\Delta$v$_{t}$} & \phantom{-}\hbox{($\sum$x$^{2}$)$^{1/2}$} \\
\hbox{} & \hbox{100\,K} & \phantom{-}\hbox{0.2\,dex} & \phantom{-}\hbox{0.2 kms$^{-1}$} & & \\
\noalign{\smallskip}
\hline
\noalign{\smallskip}
\noalign{\hrule\vskip 0.1cm}
\multicolumn{5}{c}{b14 = 2M17392719-2310311 - T$_{\rm eff}$=3643\,K, log~g=0.67} \\
\noalign{\smallskip}
\hline
\noalign{\smallskip}
P &  0.01  & 0.10 & 0.0 & 0.10  \\
S & 0.09 &  0.10   & 0.0 &  0.14  \\
K & 0.05 &   0.01  & 0.0 & 0.05 \\
\noalign{\smallskip}
\hline
\noalign{\smallskip}
\multicolumn{5}{c}{c16 = 2M17310874-2956542 - T$_{\rm eff}$=4175\,K, log~g=1.2}  \\
\noalign{\smallskip}
\hline
\noalign{\smallskip}
P &  0.02  & 0.07 & 0.0 & 0.07  \\
S & 0.04 &  0.01   & 0.0 &  0.04  \\
K & 0.05 &   0.005  & 0.0 & 0.05 \\
\noalign{\smallskip} 
\hline 
\end{tabular}
\end{flushleft}
 \end{table}

Another way to estimate uncertainties on abundances is to compare with literature. APOGEE-ASPCAP DR17 do not provide P abundances. Nevertheless, we have one star in common with \citet{brauner23}, b2 2M17173693-2806495. \citet{brauner23} reported a conservative upper limit of 1.24.
The BACCHUS Analysis of weak lines in APOGEE spectra \citep[BAWLAS;][]{hayes22,hayes23} gives the P abundance
for this same star of 0.83.
The values are different from our own determination of 0.4, but we recall that the stellar parameters are different
(calibrated vs. uncalibrated stellar parameters from APOGEE).

\begin{figure}
    \centering
    \includegraphics[width=8.5cm]{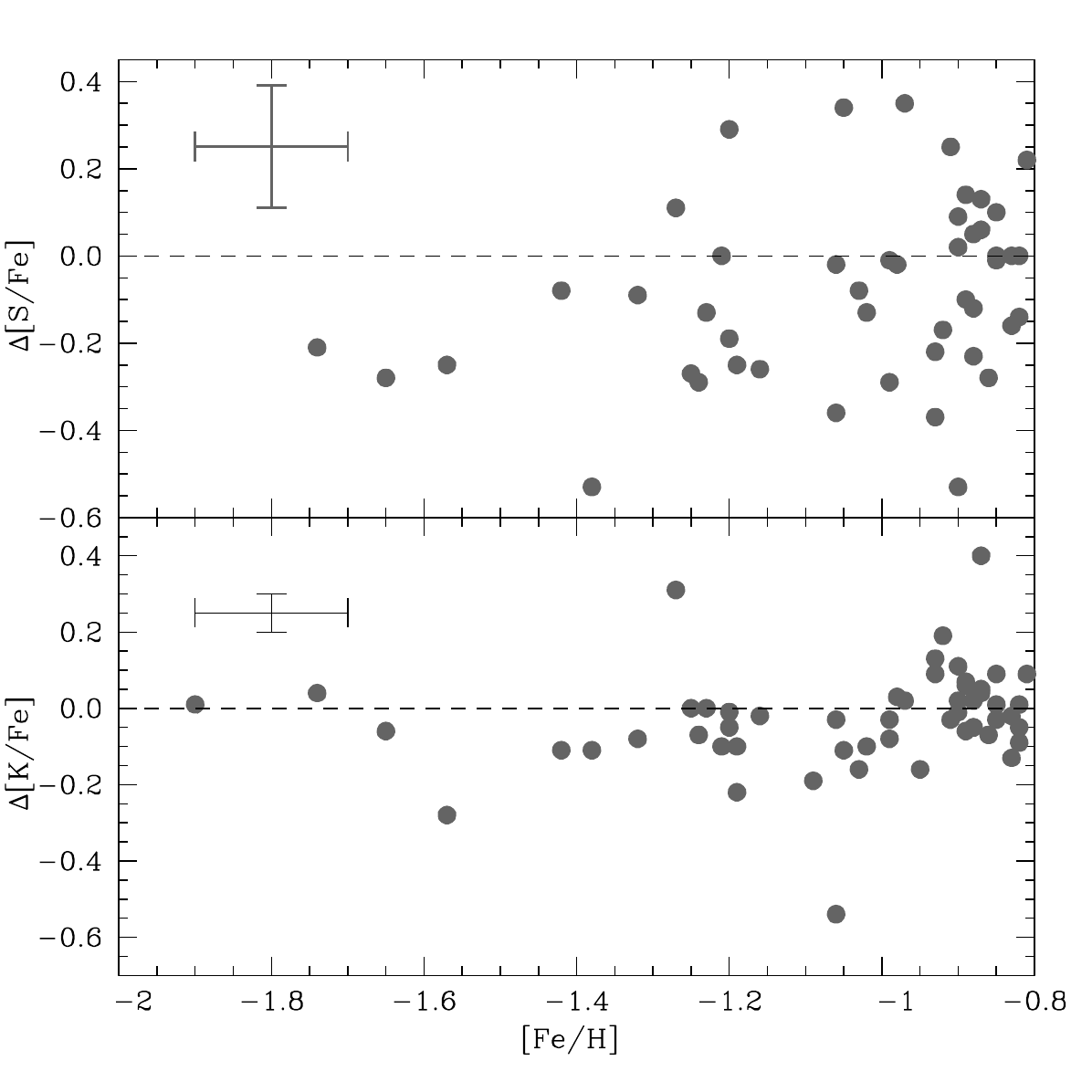}
    \caption{[S, K/Fe] vs. [Fe/H], plotting the differences between the present results and  the APOGEE DR17 values.
    Error bars correspond to the errors for the cool star b14 reported in Table \ref{errors}.}
    \label{plotdiff}
\end{figure}

Figure \ref{plotdiff} shows the difference between
the S and K abundances derived in this work minus the ones from APOGEE-ASPCAP DR17.  The mean difference between present results and ASPCAP DR17 is found to be:

\begin{equation} \label{eq1}
\begin{split}
     {\rm [S/Fe]}_{\rm present}-{\rm [S/Fe]}_{\rm ASPCAP}=-0.08^{+0.22}_{-0.17} \\
     {\rm [K/Fe]}_{\rm present}-{\rm [K/Fe]}_{\rm ASPCAP}=-0.04^{+0.07}_{-0.10}    \\
\end{split}
\end{equation}

For S the differences are due to continuum placement, but more so due to measuring only the best line for S, because, as explained above, the other lines are strongly blended with molecular lines. For K, we measure systematically lower abundances, and this also could be due to continuum placement. The fits are available under request.

\section{Present results and literature data}\label{sec:results}

The results in Table \ref{results} can be compared with literature data,
that can be essentially all described in this Section, given that there are few previous observational studies on P, S, and K.

{\it Phosphorus:}
There are only a few measurements of P abundances in the literature. 
\citet{caffau11} used the CRyogenic high-resolution InfraRed Echelle Spectrograph (CRIRES) at the Very Large Telescope (VLT) 
spectra to measure the \ion{P}{I}  10511.584,
10529.522, 10581.569, and 10596.900 {\rm \AA} lines in 20 disk dwarf stars.
\citet{caffau16} used the  GIANO spectrograph at the Telescopio Nazionale Galileo (TNG) 
to measure these same lines in another 4 disk dwarf stars.
\citet{roederer14} identified six \ion{P}{I} lines 
using the Space Telescope Imaging Spectrograph (STIS) onboard Hubble Space Telescope (HST)
in the near-UV, namely at
2154.080, 2154.121, 2533.976, 2535.603, 2553.262, and 2554.915 {\rm \AA},
relying in the end only on the \ion{P}{I} 2136 {\rm \AA} line, for 14 halo stars.
\citet{roederer16} used the Cosmic Origins spectrograph (COS) also onboard the HST,
to derive P and S abundances in carbon-rich metal-poor stars.
\citet{maas19} used the Phoenix spectrograph at the Gemini-South telescope to measure the
10581.560  and 10596.960 {\rm \AA} lines in 9 disk stars, and \citet{maas22}
used the spectra obtained with the  Hobby-Eberly Telescope at the McDonald Observatory to measure the \ion{P}{I} 10529.52 {\rm \AA} line
as the primary indicator of P abundances in 163 disk and outer halo stars.
\citet{sadakane22} used the ESPaDOnS spectrograph at the
Canada-France-Hawaii telescope (CFHT) to observe the 9750.75 and 9796.83 {\rm \AA} lines in 45 main-sequence stars.
\citet{nandakumar22} used the  Immersion GRating INfrared Spectrometer (IGRINS) at the Gemini-South telescope to measure the
\ion{P}{I} 16482.92 {\rm \AA} line for 38 disk stars.
\citet{brauner23}  analyzed 87 stars selected from the APOGEE DR17 survey, including the 16 P-rich stars previously found by \citet{masseron20}. Of this sample, 78 stars turned out to be enriched in P ([P/Fe] $>$ +0.8).

Figure \ref{plotp} shows the literature data compared with the present results.
It is interesting that a significant part of the sample stars show clearly enhanced phosphorus compared to the mean field value. P-rich stars have been previously found in other samples by \citet{masseron20}, \citet{brauner23} and \citet{brauner24}. However, we note that \citet{brauner23} found more extreme cases of P enhancement than we observe here, which led them to consider a higher limit for the P abundance ([P/Fe] $>$ +0.8) than we do here ([P/Fe] $>$ +0.45) to consider the star as P-rich. Nevertheless, it is intriguing that the P-rich stars from  \citet{masseron20,masseron20b}, \citet{brauner23}, and further discussed in \citet{brauner24} have
metallicities around [Fe/H] $\sim -$1.0, coinciding with the same
effect in our sample stars at the same metallicities. 
This might confirm our previous studies \citep{razera22,barbuy23,barbuy24}, 
namely, that there is an old population of stars that are the result of an
early fast chemical enrichment in the Milky Way towards the Galactic bulge. To confirm this conclusion,
the orbital analysis of the Masseron-Brauner sample of P-rich stars led these authors to identify most of their sample as belonging  to the thick disk
and the inner Galactic halo, thus also old populations.
Moreover, the results obtained by \citet{roederer14} seem to indicate that no P-rich stars are present among the very metal-poor halo stars ([Fe/H] $\le$ -2.0), although this sample is small and thus statistically not very robust to draw firm conclusions.

Finally, we verified possible correlations between the P abundances and the other elements C, N, O, Na, Al, Si, Ca, and iron-peak elements previously studied \citep{razera22,barbuy23,barbuy24}. 
\citet{brauner23} found a strong correlation between P and Si in  their P-rich stars sample. However, this is not as clear in our study. While it is not possible to make a direct qualitative comparison between those two studies because of  differences in stellar parameters and procedures, the difference in the correlation could be also due to the limited P abundance range of the current sample (0.5$\le$[P/Fe]$\le$1.0). Indeed, in the \citet{brauner23} within such a limited range of P abundances, the P-Si correlation tends to vanish.

{\it Sulphur:}
Figure \ref{plots} shows the present sulphur abundances compared
with literature data, as described below.
 \citet{francois87} and  \citet{rebolo01} analysed the \ion{S}{I}
8693.958, 8694.641 {\rm \AA} lines.
\citet{takada-hidai02} gathered High Resolution Echelle Spectrometer (HIRES)  observations with the samples observed by 
François (1987, 1988), and Clegg et al. (1981) for these same lines,
and derived both LTE and non-LTE abundances. We adopted their NLTE
[FeI/H] and [S/FeI] results for comparison purposes.
\citet{chen02} used these same lines
plus the lines \ion{S}{I} 6046.03, 6052.67, 6757.17 {\rm \AA}.
\citet{takada-hidai02} used the 8693.958 and 8694.641 {\rm \AA} lines
to derive the S abundances of giants and dwarfs from
several samples, including their own, plus
those from \citet{francois87,francois88}, and \citet{clegg81}.
\citet{korn05} used these lines and the \ion{S}{I}   9212.9, 9228.1
and 9237.5 {\rm \AA} lines.
\citet{nissen07} analysed the weaker \ion{S}{I}  8694.6 and 
the stronger \ion{S}{I} 9212.9 and 9237.5 {\rm \AA} lines, and
presented abundances derived in LTE and NLTE.
\citet{spite11} used several lines, namely, \ion{S}{I} 
6756.851, 6757.007, 6757.171, 8693.931, 8694.626, 9212.863, 9228.093,
9237.538, 10455.449, 10456.757, 10459.406, and 10821.176  {\rm \AA} lines,
and carried out NLTE calculations.
 \citet{caffau16}
 measured the 10455.449,  10456.757, and  10459.406 {\rm \AA} lines,
 and computed NLTE abundances.
\citet{perdigon21} analysed the  \ion{S}{I} 6743.440, 6743.531,
6743.640, 6748.573, 6748.682, 6748.837, 6756.851, 6757.007, 6757.171
 {\rm \AA} lines, and computed LTE abundances, arguing that NLTE
 corrections would be smaller than 0.1\,dex following
 \citet{takeda16} and \citet{korotin17}.
 \citet{lucertini22} measured the lines \ion{S}{I}   6757.171, 8694.626,
 9212.863, 9228.093, and 9237.538 {\rm \AA}.

{\it Potassium:}
Figure \ref{plotk} shows the K abundances for our sample of 58
stars compared with literature data.
\citet{zhang06} derived K abundances from the \ion{K}{I}
7664.92, and 7698.98 {\rm \AA} lines
for moderately metal-poor disk stars, providing results corrected
for NLTE effects.
\citet{andrievsky10} computed NLTE K abundances for the sample halo
stars from \citet{cayrel04}, for the lines \ion{K}{I}
5801.752,  7698.965, and 12432.274 {\rm \AA}.
\citet{reinhard24} derived LTE K abundances for 20 halo stars.

\section{Chemical-evolution of P, S, K}\label{models}

\subsection{Nucleosynthesis}

{\it Phosphorus:} \citet{brauner24} considered the nucleosynthesis options to explain
the P excess in stars of metallicity [Fe/H] $\approx -$1.0,
and could not identify a process responsible for this effect.
Normally, P is produced in CCSNe through neutron captures on the Si isotopes.
$^{31}$P, together with
$^{28}$Si and $^{30}$Si, are produced in the oxygen- and neon-burning
shells, and then ejected at the supernova core-collapse event 
 \citep{woosley95}. During the explosion, 
according to \citet{pignatari16}, there is some
production of  $^{31}$P by explosive C burning and explosive He
burning.
One possible explanation for the large excess of P was given by
 \citet{bekki24}  through P production in oxygen-neon (ONe) novae.
 However, it is not clear why this would take place only in the environment
 where the old bulge stars of moderate metallicity were formed.

There is also the possibility of enrichment by Asymptotic Giant Branch
(AGB) stars, where  $^{31}$P is the result of neutron-capture on $^{30}$Si. However, this latter contribution is less likely as this is expected to occur in only very metal-poor AGBs \citep{karakas16}.

{\it Sulphur:} Sulphur is an $\alpha$-element, therefore it is expected
to behave like O, Mg, Si, and Ca, being produced in
massive stars and ejected by SNe II.

{\it Potassium:} $^{39}$K is the majority isotope, comprising 93.132\%, followed by $^{41}$K with
6.721\% and $^{40}$K with 0.147\% \citep{asplund09}. The $^{39,41}$K isotopes are dominantly produced
in CCSNe \citep{pignatari16}.

\subsection{Chemical-evolution models}

The chemical-evolution model for the Galactic bulge is derived from the
chemical-evolution model for elliptical galaxies of \citet{friaca98}. This model consists of a multi-zone chemical evolution
coupled with a hydrodynamical code. For the Galactic bulge, a
classical spheroid with a baryonic mass of 2$\times$10$^{9}$ M$_{\odot}$
and a dark halo mass of 1.3$\times$10$^{10}$ M$_{\odot}$ are assumed.
Cosmological parameters from the \citet{planck20} are adopted, namely
$\Omega_{m}$ = 0.31, $\Omega_{\Lambda}$ = 0.69, 
Hubble constant H$_{0}$= 68 km s$^{-1}$Mpc$^{-1}$,
and an age of the Universe of 13.801$\pm$0.024 Gyr.

For the nucleosynthesis yields, we adopt:
(i) for massive stars,
the metallicity-dependent yields from CCSNe/SNe II from \citet{woosley95}, 
and for low metallicities (Z $<$ 0.01 Z$_{\odot}$ , or [Fe/H] $< -$2.5, the yields
are from high-explosion-energy hypernovae (HNe) from \citet{nomoto13};
(ii) Type Ia Supernovae (SNIa) yields are from \citet{iwamoto99} – their models W7 
(progenitor star of initial metallicity
Z = Z$_{\odot}$) and W70 (zero initial metallicity); and (iii) for intermediate-mass stars (0.8–8 M$_{\odot}$) with initial Z = 0.001, 0.004, 0.008, 0.02, and
0.4, we adopt yields from \citet{vandenhoek97}  with variable $\eta$ (AGB case).

Our nucleosynthesis prescriptions  also consider neutrino-interaction processes, which could be important at very low metallicities. Core-collapse supernovae release large amounts of energy as neutrinos ($>$ 10$^{53}$ erg) during the formation of the neutron star or black hole. In this case, the interaction of these neutrinos with matter could significantly increase the yields of odd-Z elements \citep{yoshida08}. Therefore, for the supernovae with very 
low-metallicity progenitors, our models include the enhancements by neutrino processes of the production of odd-Z elements such as phosphorus and potassium. In this paper, we adopt the results of the calculations of \citet{yoshida08} for the case with the total neutrino energy of the supernova explosion E$_{\nu}=3\times 10^{53}$ erg and a neutrino temperature T$_{\nu}$ = 8 Mev.

In our calculations, we multiplied the yields from WW95 for subsolar metallicities by a factor 2 for P, and a factor 3 for K. The factor 2 for P is the same factor adopted for cobalt, as suggested by \citet{timmes95} and adopted by \citet{ernandes20}. The value of this correction comes from the comparison of [P/Fe] vs. [Fe/H] given by Figure 23 of  
\citet{timmes95}  and the observations of \citet{caffau11}. In the case of K, the Galactic chemical evolution model of \citet{timmes95}, using yields of WW95, shows a sharp drop of [K/Fe] below [Fe/H] $ \approx$ 0.6 of in serious disagreement with the observations (see their figure 24). The factor 2 for P is similar, but smaller than the factor of 2.75 adopted by \citet{cescutti12}.
These corrections reflect the perception that the WW95 yields for Z $<$ Z${\odot}$ tend to underestimate the production of some odd-Z elements. On the other hand, for [Fe/H] $<$ $-$2.5, our models, that include hypernovae and neutrino-process in supernovae, account for the observations and do not need further corrections.

Models are computed for radii of r $< 0.5$, $0.5 <$ r $< 1$, $1 <$ r $< 2$,
and $2 <$ r $< 3$ kpc from the Galactic centre, and for specific star-formation rate
values of $\nu = 1$ and $3$ Gyr$^{-1}$.

\begin{figure}
    \centering
    \includegraphics[width=8.5cm]{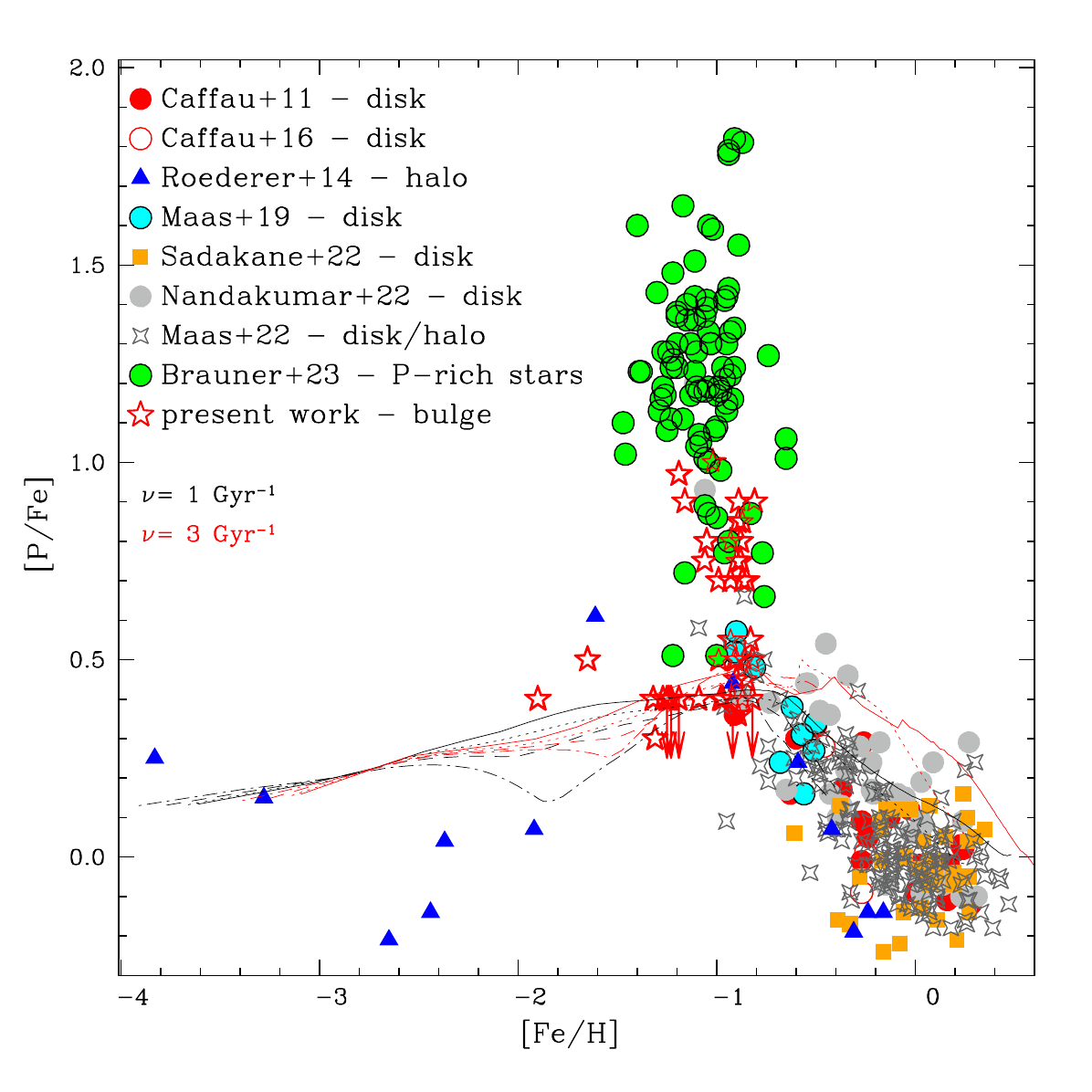}
    \caption{[P/Fe] vs. [Fe/H] for the present results compared with literature data.
    Symbols -- red-open stars: present work, 
    red-filled circles: \citet{caffau11}, red- open circles: \citet{caffau16},
    blue-filled triangles: \citet{roederer14}, filled-cyan circles + open-black circles:
    \citet{maas19}, light grey-filled circles: \citet{nandakumar22}, 
    light grey open 4-side stars: \citet{maas22}, green-filled circles + 
    open-black circles: \citet{brauner23}.
    Different model lines correspond to the outputs of models computed for radii r $<$ 0.5, 0.5 $<$ r $<$ 1, 1 $<$ r $<$ 2, 
    and 2 $<$ r $<$ 3 kpc from the Galactic centre. Black lines correspond to
    specific star formation $\nu$ = 1 Gyr$^{-1}$, red lines to $\nu$ = 3 Gyr$^{-1}$.}
    \label{plotp}
\end{figure}

Figure \ref{plotp} over-plots the chemical-evolution models for specific star-formation rates
of $\nu$ = 1 and 3 Gyr$^{-1}$ for the phosphorus data. It is interesting  to note that, at
metallicities [Fe/H] $\sim -$1.0, there is a maximum  [P/Fe] = +0.423 at [Fe/H] = $-$0.825 for the model with $\nu$ = 1, and
[P/Fe] = +0.479 at Fe/H] = $-$0.827 for the model with $\nu$ = 3.
Moreover, the maximum [P/Fe] in the models approximately coincide with
the metallicities of the P-rich stars, although the very high levels of P are not reached.
This maximum arises from the yields of the \citet{woosley95} models.
About one third of the stars show higher values relative to the models.
In summary, the nucleosynthesis scenario for P-excess has not been identified yet, and different possibilities remain open.

\begin{figure}
    \centering
    \includegraphics[width=8.5cm]{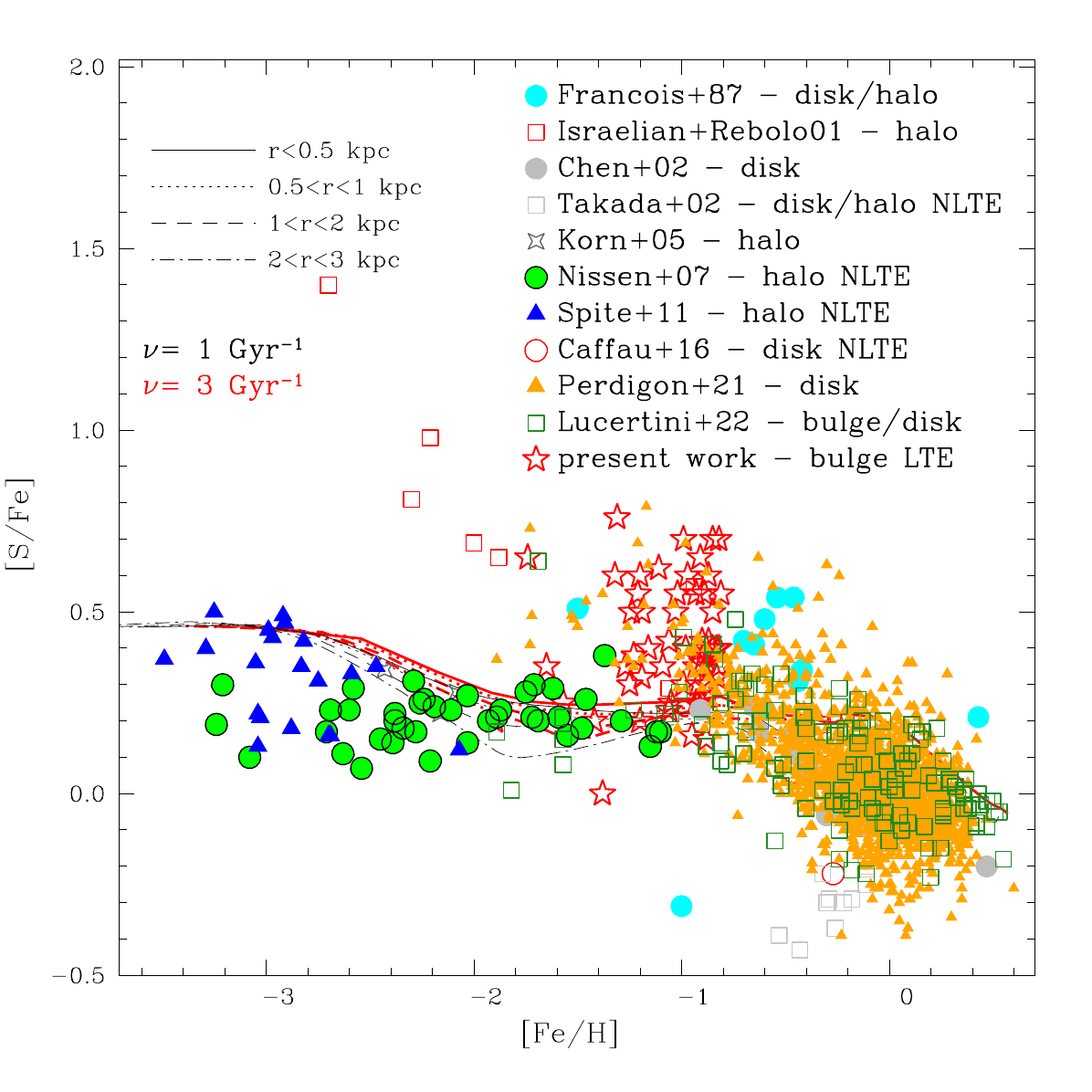}
    \caption{[S/Fe] vs. [Fe/H] for the present results compared with literature data
    Symbols -- red-open stars: present work,
    cyan-filled circles: \citet{francois87}, red-open squares: \citet{rebolo01}, 
    light grey- filled circles: \citet{chen02}, light-grey 4-side stars: \citet{korn05},
    green- filled circles + black- open circles: \citet{nissen07}, 
    blue- filled triangles: \citet{spite11}, red- open circles: \citet{caffau16}, 
    orange- filled triangles: \citet{perdigon21}, forest green- open squares: \citet{lucertini22}.
    Same as in Fig. \ref{plotp} for the model lines.
 }
    \label{plots}
\end{figure}

Figure \ref{plots} over-plots the chemical-evolution models for specific star-formation rates of $\nu$ = 1 and 3 Gyr$^{-1}$, for our results and literature data for sulphur described above. Our data seems to indicate that sulphur is enhanced in the bulge. Although a first thought is that  NLTE effects could be artificially enhancing the abundance and the spread,
the grid of NLTE corrections by \citet{korotin25} indicates only rather low values for the corrections.
On the other hand, 
the amplitude of such effect on sulphur can be notably illustrated by comparing the LTE studies \citep[e.g.][]{rebolo01} - who also found enhanced sulphur in halo stars- to other studies of similar population but in NLTE \citep[e.g. ]{nissen07}. 
We also argue that, sulphur as an $\alpha$-element is not expected to behave distinctively from other $\alpha$-elements such as O, Mg, Ca.  
Previous studies of bulge stars \citep{gonzalez2011} have demonstrated that  Mg, Ca or Ti abundances in metal-poor bulge stars are indistinguishable from those of the thick disk and halo stars at the same metallicities.
Therefore, we conclude that our sulphur abundances are very likely overestimated, 
and that S abundances are expected to be  compatible with the ones of the disk and halo at similar metallicities such as found in other stars of the bulge by \citet[e.g.][]{lucertini22}. 

Regarding the comparison of the data with the models, we confirm that the latter satisfactorily reproduce the behaviour of [S/Fe] vs. [Fe/H] over the full metallicity range.

\begin{figure}
    \centering
    \includegraphics[width=8.5cm]{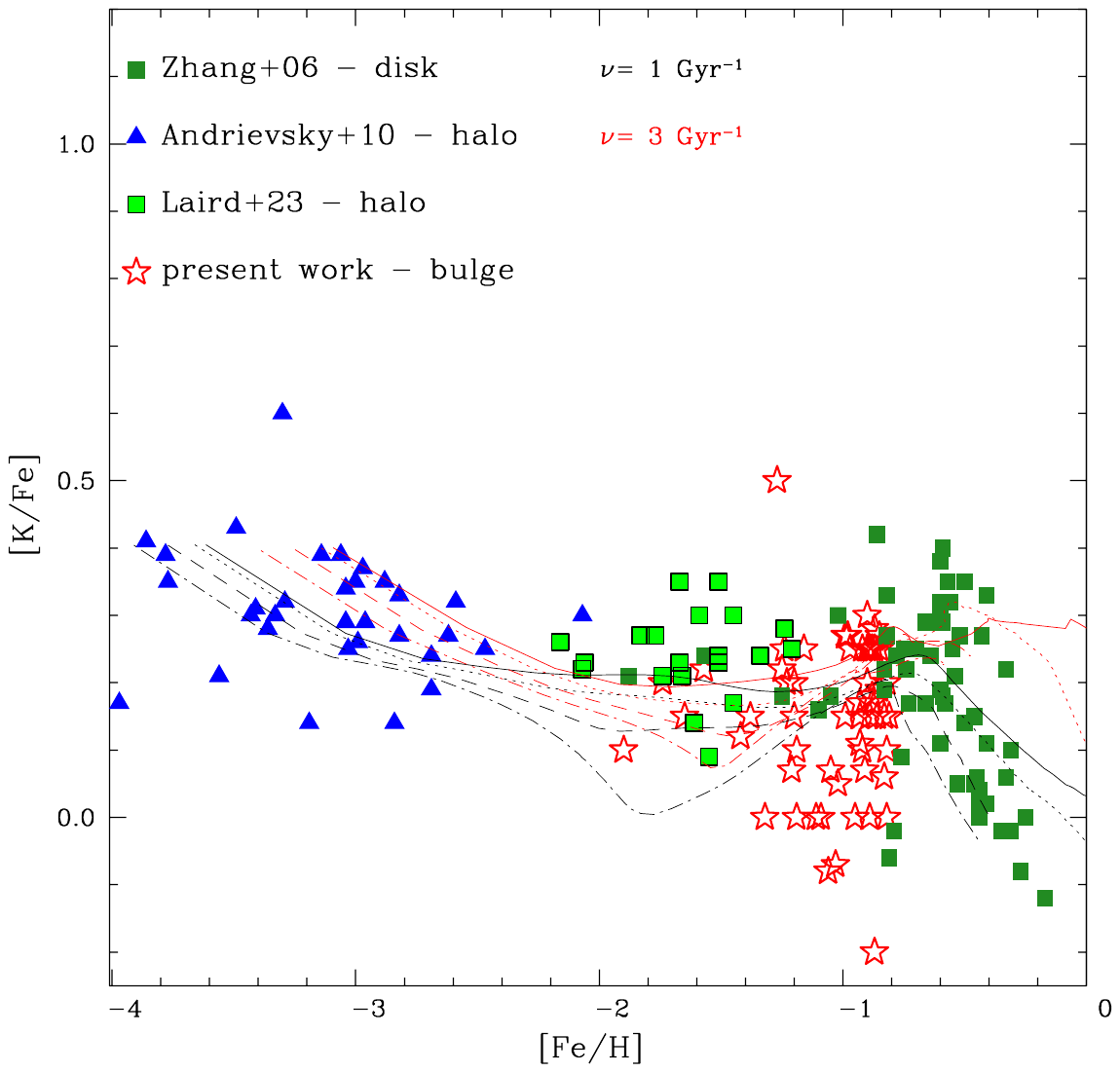}
    \caption{[K/Fe] vs. [Fe/H] for the present results compared with literature data
    Symbols -- red- open stars: present work,
    dark- green squares: \citet{zhang06}, blue-filled triangles: \citet{andrievsky10}, 
    green- filled squares + black- open squares: \citet{reinhard24}.
    Same as in Fig. \ref{plotp} for the model lines.}
    \label{plotk}
\end{figure}

Figure \ref{plotk} over-plots the results for K, with the models showing very good agreement with the data.

\section{Conclusions}\label{conclusions}

This is the fourth paper of a series dealing with the chemical-abundance analysis of 58 stars selected to have 
[Fe/H] $<-$0.8 and orbits compatible with being members of a spheroidal component of the Galactic bulge. In the present work, we used suitable lines to compute the abundances of the  elements P, S, and K. 
The abundances of P were not available from the
APOGEE-ASPCAP DR17 results.

A striking result are the relative over-abundances of P for about one third of the sample stars, 
a phenomenon that may be related to the P-rich stars phenomenon by  \citet{masseron20,masseron20b}, \citet{brauner23}
and \citet{brauner24} although more moderate. 
In the present work, the over-abundance of P is seen in part of these old bulge stars, for the first time, and from literature data.
For the other two thirds of the sample the P abundances are compatible with the chemical evolution models.
Further samples should be analysed in order to have a more solid statistics on the number of P-rich vs. P-normal stars in the Galactic bulge.
It is remarkable that no P-rich stars are  present in  the most metal-poor bulge stars as well as in the very metal-poor halo stars. It appears that
these over-abundances are not observed in other stellar populations, and might indicate the contribution of a particular kind of early supernovae operating in the central part of the Galaxy. 
The nucleosynthesis process involved is, however not yet clearly identified. 

 Sulphur is an $\alpha$-element, and behaves as such, as seen in the comparison with chemical-evolution models. 
 We stress though that the present results rely on one line, and the abundances have not been corrected for non-LTE departures, 
 given that the \citet{korotin25} grid is limited to higher temperatures relative to our sample. 
 Their corrections indicate a negligible correction,
 but this problem should be further investigated, in order to explain the discrepancies between our results and the models.

Potassium is another little-studied element. The chemical-evolution models fit the behaviour of [K/Fe] vs. metallicity
 remarkably well; this is partly due to the inclusion of the $\nu$-process yields in the models.

\begin{acknowledgements}
B.B. and A.C.S.F. acknowledge grants from FAPESP, Conselho Nacional de Desenvolvimento Cient\'ifico e Tecnol\'ogico (CNPq) and Coordena\c{c}\~ao de Aperfei\c{c}oamento de Pessoal de N\'ivel Superior (CAPES) - Financial code 001. H.E. acknowledges a post-doctoral fellowship at Lund Observatory.
M.S.C. acknowledges a CAPES doctoral fellowship.
P.S. acknowledges the FAPESP post-doctoral fellowships 2020/13239-5 and 2022/14382-1.
S.O.S. acknowledges the support from Dr. Nadine Neumayer's Lise Meitner grant from the Max Planck Society.
A.P.-V., B.B., and S.O.S. acknowledge the DGAPA-PAPIIT grant IA103224.
BB, HE, PS and SOS are part of the Brazilian Participation Group (BPG) in the Sloan Digital Sky Survey (SDSS), from the
Laborat\'orio Interinstitucional de e-Astronomia – LIneA, Brazil.
T.M. acknowledges support from the Spanish Ministry of Science and Innovation with the the proyecto plan nacional \textit{PLAtoSOnG} (grant no. PID2023-146453NB-100).
M.B. acknowledges financial support from the European Union and the State Agency of Investigation of the Spanish Ministry of Science and Innovation (MICINN) under the grant PRE-2020-095531 of the Severo Ochoa Program for the Training of Pre-Doc Researchers (FPI-SO).
J.G.F-T gratefully acknowledges the grant support provided by Proyecto Fondecyt Iniciaci\'on No. 11220340, 
Proyecto Fondecyt Postdoc No. 3230001 (Sponsored by J.G.F-T)  and from the Joint Committee ESO-Government of Chile 2021 (ORP 023/2021), and 2023 (ORP 062/2023). 
F.A. acknowledges partial supported by the Spanish MICIN/AEI/10.13039/501100011033 and by "ERDF A way of making Europe" by the “European Union” through grant PID2021-122842OB-C21, and the Institute of Cosmos Sciences University of Barcelona (ICCUB, Unidad de Excelencia ’Mar\'{\i}a de Maeztu’) through grant CEX2019-000918-M. FA acknowledges the grant RYC2021-031683-I funded by MCIN/AEI/10.13039/501100011033 and by the European Union NextGenerationEU/PRTR.
D.M. gratefully acknowledges support from the Center for Astrophysics and Associated Technologies (CATA) by ANID BASAL projects ACE210002 and FB210003, and Fondecyt Project No. 1220724.
D.G. gratefully acknowledges the support provided by Fondecyt regular n. 1220264.
D.G. also acknowledges financial support from the Direcci\'on de Investigaci\'on y Desarrollo de la Universidad de La Serena through the Programa de Incentivo a la Investigaci\'on de Acad\'emicos (PIA-DIDULS).
The work of V.V.S. and V.M.P. is supported by NOIRLab, which is managed by the Association of Universities for Research in Astronomy (AURA) under a cooperative agreement with the U.S. National Science Foundation.
T.C.B. acknowledges support from grant PHY 14-30152; Physics Frontier Center/JINA Center for the Evolution of the Elements (JINA-CEE), and from OISE-1927130: The International Research  Network for Nuclear Astrophysics (IReNA), awarded by the US National Science Foundation.
Apogee project: Funding for the Sloan Digital Sky Survey IV has been provided by the Alfred P. Sloan Foundation, the U.S. Department of Energy Office of Science, and the Participating Institutions. SDSS acknowledges support and resources from the Center for High-Performance Computing at the University of Utah. The SDSS web site is www.sdss.org. 
SDSS is managed by the Astrophysical Research Consortium for the Participating Institutions of the SDSS Collaboration including the Brazilian Participation Group, the Carnegie Institution for Science, Carnegie Mellon University, Center for Astrophysics | Harvard \& Smithsonian (CfA), the Chilean Participation Group, the French Participation Group, Instituto de Astrof\'isica de Canarias, The Johns Hopkins University, Kavli Institute for the Physics and Mathematics of the Universe (IPMU) / University of Tokyo, the Korean Participation Group, Lawrence Berkeley National Laboratory, Leibniz Institut f\"ur Astrophysik Potsdam (AIP), Max-Planck-Institut f\"ur Astronomie (MPIA Heidelberg), Max-Planck-Institut f\"ur Astrophysik (MPA Garching), Max-Planck-Institut f\"ur Extraterrestrische Physik (MPE), National Astronomical Observatories of China, New Mexico State University, New York University, University of Notre Dame, Observat\'orio Nacional / MCTI, The Ohio State University, Pennsylvania State University, Shanghai Astronomical Observatory, United Kingdom Participation Group, Universidad Nacional Aut\'onoma de M\'exico, University of Arizona, University of Colorado Boulder, University of Oxford, University of Portsmouth, University of Utah, University of Virginia, University of Washington, University of Wisconsin, Vanderbilt University, and Yale University.
\end{acknowledgements}



\bibliographystyle{aa} 
\bibliography{bibliogpsk}



\begin{appendix}


\begin{table*}[!ht]
\centering
\caption{Internal and 2MASS star identification, the ASPCAP uncalibrated stellar parameters, recomputed C, N, O abundances, the abundances of P, S, and K derived in this work, and
the APOGEE-ASPCAP abundances of S and K.} 
\resizebox{0.9\textwidth}{!}{
\begin{tabular}{cccccccccccccccccccccccccc}
\noalign{\smallskip}
\hline
\noalign{\smallskip}
\hbox{ID} & \hbox{ID}&
T$_{\rm eff}$ &\hbox{log~g} & \hbox{[Fe/H]} &  \hbox{v$_t$} &  [C/Fe] & [N/Fe] & [O/Fe] & [P/Fe] &  [S/Fe] & [S/Fe]$_{APO}$ & [K/Fe] & [K/Fe]$_{APO}$ & \\
\hbox{(internal)} & \hbox{(2MASS)} & \hbox{(K)} & &  &  \hbox{(km/s)} & & & & & & & & &  \\  
 \noalign{\smallskip}
\noalign{\hrule}
\hline                   
b1 &2M17153858-2759467 & 3922.7$\pm$10 & 0.34$\pm$0.05 &  $-$1.65$\pm$0.02 & 2.62 &b1 $-$0.60$\pm$0.03 & +0.25$\pm$0.03 & +0.20$\pm$0.03 & +0.50$\pm$0.10   & +0.35$\pm$0.05 &  +0.63/{\bf 0.41}$\pm$0.08  &  +0.15$\pm$0.03    & +0.21$\pm$0.09& \\   
b2 &2M17173693-2806495 & 3908.9$\pm$8 &  0.95$\pm$0.04 &  $-$0.97$\pm$0.01 & 2.20 &b2 +0.15$\pm$0.03 & +0.10$\pm$0.03 & +0.38$\pm$0.03   & +0.40$\pm$0.05   & +0.60$\pm$0.05 &  +0.25$\pm$0.09             &  +0.25$\pm$0.03    & +0.23$\pm$0.06& \\   
b3 &2M17250290-2800385 & 3796.6$\pm$6 &  0.91$\pm$0.03 &  $-$0.82$\pm$0.01 & 2.39 &b3 +0.15$\pm$0.03 & +0.10$\pm$0.03 & +0.40$\pm$0.03   & +0.50$\pm$0.05   & +0.40$\pm$0.05 &  +0.54/{\bf 0.52}$\pm$0.06  &  +0.00$\pm$0.03    & +0.09$\pm$0.05& \\   
b4 &2M17265563-2813558 & 4096.2$\pm$14 & 1.00$\pm$0.06 &  $-$1.32$\pm$0.02 & 1.89 &b4 $-$0.30$\pm$0.03 & +0.30$\pm$0.08 & +0.20$\pm$0.03 & +0.40$\pm$0.10   & +0.60$\pm$0.05 &  +0.69$\pm$0.19             &  +0.00$\pm$0.08    & +0.08$\pm$0.10& \\   
b5 &2M17281191-2831393 & 4029.1$\pm$7 &  0.96$\pm$0.04 &  $-$1.19$\pm$0.01 & 1.73 &b5 $-$0.30$\pm$0.03 & +0.30$\pm$0.03 & +0.40$\pm$0.03 &$<$+0.40          & +0.50$\pm$0.05 &  +0.75/{\bf 0.58}$\pm$0.01  &  +0.00$\pm$0.03    & +0.22$\pm$0.06& \\   
b6 &2M17295481-2051262 & 4205.9$\pm$13 & 1.50$\pm$0.05 &  $-$0.85$\pm$0.02 & 1.71 &b6 +0.15$\pm$0.10 & +0.20$\pm$0.10 & +0.40$\pm$0.05    &    ---          & +0.40$\pm$0.05 &  +0.30$\pm$0.13             &  +0.19$\pm$0.03    & +0.10$\pm$0.08& \\   
b7 &2M17303581-2354453 & 3863.0$\pm$9 &  0.77$\pm$0.04 &  $-$0.98$\pm$0.02 & 2.13 &b7 +0.07$\pm$0.03 & +0.05$\pm$0.08 & +0.50$\pm$0.03    & +0.40$\pm$0.10  & +0.40$\pm$0.05 &  +0.42$\pm$0.12             &  +0.27$\pm$0.03    & +0.24$\pm$0.07& \\   
b8 &2M17324257-2301417 & 3668.2$\pm$7 &  0.79$\pm$0.04 & $-$0.82$\pm$0.02 & 2.30 &b8 -0.10$\pm$0.03 & +0.15$\pm$0.05 & +0.35$\pm$0.05    &$<$+0.40         & +0.70$\pm$0.05 &  {\bf 0.61}$\pm$0.07        &  +0.20$\pm$0.03    & +0.19$\pm$0.06& \\   
b9 &2M17330695-2302130 & 3566.6$\pm$6 &  0.35$\pm$0.03 &  $-$0.93$\pm$0.01 & 2.42 &b9 +0.30$\pm$0.03 & +0.00$\pm$0.03 & +0.55$\pm$0.03    & +0.80$\pm$0.10  & +0.55$\pm$0.05 &  ---                        &  +0.15$\pm$0.03    &  ---          & \\   
b10 &2M17344841-4540171& 3869.2$\pm$6 &  0.85$\pm$0.03 &  $-$0.88$\pm$0.01 & 2.16 &b10 $-$0.30$\pm$0.03 & +0.20$\pm$0.03 & +0.35$\pm$0.03 & +0.80$\pm$0.10  & +0.15$\pm$0.10 &  +0.27$\pm$0.07             &  +0.18$\pm$0.03    & +0.23$\pm$0.05& \\   
b11 &2M17351981-1948329& 3553.5$\pm$5 &  0.44$\pm$0.03 &  $-$1.11$\pm$0.01 & 3.06 &b11 $-$0.10$\pm$0.03 & +0.20$\pm$0.03 & +0.30$\pm$0.05 &     ---         & +0.62$\pm$0.05 &  ---                        &  +0.00$\pm$0.03    &  ---          & \\   
b12 &2M17354093-1716200& 3895.5$\pm$7 &  1.01$\pm$0.03 &  $-$0.87$\pm$0.01 & 2.02 &b12 +0.05$\pm$0.03 & +0.30$\pm$0.03 & +0.50$\pm$0.03   & +0.70$\pm$0.10  & +0.42$\pm$0.05 &  +0.29$\pm$0.08             &  +0.28$\pm$0.03    & +0.24$\pm$0.05& \\   
b13 &2M17390801-2331379& 3740.4$\pm$6 &  0.83$\pm$0.03 &  $-$0.81$\pm$0.01 & 2.35 &b13 +0.05$\pm$0.03 & +0.20$\pm$0.03 & +0.40$\pm$0.03   & +0.90$\pm$0.10$^{*}$& +0.55$\pm$0.05 &  +0.33/{\bf 0.66}$\pm$0.09  &  +0.15$\pm$0.08    & +0.06$\pm$0.05& \\ 
b14 &2M17392719-2310311& 3643.3$\pm$6 &  0.67$\pm$0.04 &  $-$0.87$\pm$0.01 & 2.55 &b14 +0.20$\pm$0.03 & +0.40$\pm$0.03 & +0.60$\pm$0.03   & +0.40$\pm$0.05  & +0.38$\pm$0.05 &  ---                        &  +0.25$\pm$0.03    & +0.20$\pm$0.06& \\   
b15 &2M17473299-2258254& 4018.3$\pm$9 &  0.47$\pm$0.05 &  $-$1.74$\pm$0.01 & 2.12 &b15 $-$0.70$\pm$0.03 & +0.80$\pm$0.03 & +0.35$\pm$0.03 & ---             & +0.65$\pm$0.05 &  +0.86/{\bf 0.63}$\pm$0.03  &  +0.20$\pm$0.10    & +0.16$\pm$0.08& \\   
b16 &2M17482995-2305299& 4213.6$\pm$9 &  1.24$\pm$0.04 &  $-$1.03$\pm$0.01 & 2.10 &b16 $-$0.30$\pm$0.10 & +0.30$\pm$0.10 & +0.30$\pm$0.08 &   ---           & +0.50$\pm$0.05 &  +0.58$\pm$0.09             &$-$0.07$\pm$0.03    & +0.09$\pm$0.06& \\   
b17 &2M17483633-2242483& 3651.5$\pm$6 &  0.44$\pm$0.03 &  $-$1.09$\pm$0.01 & 2.58 &b17 $-$0.23$\pm$0.03 & +0.10$\pm$0.03 & +0.35$\pm$0.03 & +0.40$\pm$0.05  & +0.35$\pm$0.05 &  ---                        &  +0.00$\pm$0.03    & +0.19$\pm$0.05& \\   
b18 &2M17503263-3654102& 3893.5$\pm$6 &  0.64$\pm$0.03 &  $-$0.99$\pm$0.01 & 2.19 &b18 $-$0.10$\pm$0.03 & +0.30$\pm$0.03 & +0.25$\pm$0.05 & +0.70$\pm$0.05  & +0.30$\pm$0.05 &  +0.59$\pm$0.07             &  +0.27$\pm$0.03    & +0.30$\pm$0.05& \\   
b19 &2M17552744-3228019& 4018.9$\pm$8 &  1.00$\pm$0.04 &  $-$1.06$\pm$0.01 & 2.00 &b19 $-$0.40$\pm$0.03 & +0.10$\pm$0.05 & +0.10$\pm$0.08 &    ---          & +0.23$\pm$0.05 &  +0.59/{\bf 0.46}$\pm$0.08  &$-$0.33$\pm$0.03    & +0.21$\pm$0.06& \\   
b20 &2M18020063-1814495& 3988.8$\pm$10 & 0.80$\pm$0.05 &  $-$1.38$\pm$0.02 & 2.04 &b20 $-$0.30$\pm$0.05 & $-$0.10$\pm$0.05 & +0.00$\pm$0.05 &    ---        & +0.00$\pm$0.10 &  +0.53$\pm$0.15             &  +0.15$\pm$0.03    & +0.26$\pm$0.08& \\   
b21 &2M18050452-3249149& 3940.8$\pm$7 &  0.77$\pm$0.04 &  $-$1.16$\pm$0.01 & 2.08 &b21 $-$0.40$\pm$0.03 & +0.20$\pm$0.03 & +0.40$\pm$0.03  & +0.90$\pm$0.10 & +0.40$\pm$0.05 &  +0.66$\pm$0.08             &  +0.25$\pm$0.03    & +0.27$\pm$0.06& \\   
b22 &2M18050663-3005419& 3439.9$\pm$5 &  0.23$\pm$0.03 &  $-$0.92$\pm$0.01 & 2.52 &b22 +0.12$\pm$0.03 & +0.20$\pm$0.03 & +0.35$\pm$0.03    & $<$0.40     & +0.30$\pm$0.05 &  ---                        &  +0.10$\pm$0.03    & ---           & \\   
b23 &2M18065321-2524392& 3893.1$\pm$7 &  0.95$\pm$0.04 &  $-$0.89$\pm$0.01 & 2.02 &b23 +0.00$\pm$0.03 & +0.20$\pm$0.03 & +0.38$\pm$0.03    & +0.90$\pm$0.05 & +0.38$\pm$0.05 &  +0.24$\pm$0.07             &  +0.00$\pm$0.03    & +0.06$\pm$0.05& \\   
b24 &2M18104496-2719514& 4153.1$\pm$11 & 1.33$\pm$0.04 &  $-$0.82$\pm$0.02 & 2.05 &b24 $-$0.10$\pm$0.10 & $-$0.10$\pm$0.10 & +0.10$\pm$0.10 &    ---        & +0.35$\pm$0.10 &  +0.35$\pm$0.11             &  +0.10$\pm$0.03    & +0.15$\pm$0.07& \\   
b25 &2M18125718-2732215& 3617.2$\pm$6 &  0.44$\pm$0.03 &  $-$1.31$\pm$0.01 & 2.64 &b25 +0.00$\pm$0.03 & +0.40$\pm$0.03 & +0.45$\pm$0.03    & +0.30$\pm$0.05 & +0.76$\pm$0.05 &  {\bf 0.88}$\pm$0.05         &   ---              &$-$0.85$\pm$0.06& \\  
b26 &2M18200365-3224168& 3976.6$\pm$6 &  0.95$\pm$0.04 &  $-$0.86$\pm$0.01 & 1.94 &b26 $-$0.20$\pm$0.08 & +0.20$\pm$0.08 & +0.32$\pm$0.03  & +0.50$\pm$0.05 & +0.30$\pm$0.05 &  +0.58/{\bf 0.41}$\pm$0.05  &  +0.15$\pm$0.03    & +0.22$\pm$0.05& \\    
b27 &2M18500307-1427291& 4076.0$\pm$9 &  1.23$\pm$0.04 &  $-$0.95$\pm$0.01 & 1.73 &b27 $-$0.10$\pm$0.03 & +0.20$\pm$0.03 & +0.30$\pm$0.08  &    ---         & +0.16$\pm$0.05 &  +0.80$\pm$0.09             &  +0.00$\pm$0.03    & +0.16$\pm$0.06& \\    
c1 &2M17173248-2518529 & 3977.0$\pm$9 &  1.00$\pm$0.04 &  $-$0.91$\pm$0.02 & 1.81 &c1 $-$0.10$\pm$0.03 & +0.20$\pm$0.03 & +0.33$\pm$0.03   & +0.50$\pm$0.10 & +0.65$\pm$0.05 &  +0.40$\pm$0.11             &  +0.07$\pm$0.03    & +0.10$\pm$0.07& \\    
c2 &2M17285088-2855427 & 3838.0$\pm$0 &  0.63$\pm$0.04 &  $-$1.23$\pm$0.02 & 2.18 &c2  $-$0.30$\pm$0.03 & +0.20$\pm$0.03 & +0.50$\pm$0.03  & $<$+0.40 & +0.38$\pm$0.05 &  +0.51$\pm$0.13             &  +0.20$\pm$0.03    & +0.20$\pm$0.07& \\    
c3 &2M17301495-2337002 & 3814.0$\pm$6 &  0.69$\pm$0.03 &  $-$1.06$\pm$0.01 & 2.22 &c3 $-$0.35$\pm$0.03 & +0.00$\pm$0.03 & +0.25$\pm$0.03   & +0.75$\pm$0.05 & +0.42$\pm$0.05 &  +0.44$\pm$0.08             &  $-$0.08$\pm$0.03  & $-$0.05$\pm$0.05 \\   
c4 &2M17453659-2309130 & 4133.1$\pm$9 &  1.27$\pm$0.04 &  $-$1.20$\pm$0.01 & 1.08 &c4 $-$0.30$\pm$0.05 & +0.10$\pm$0.08 & +0.25$\pm$0.05   &   ---          & +0.20$\pm$0.05 &  +0.39$\pm$0.11             &  +0.15$\pm$0.03    & +0.20$\pm$0.07& \\    
c5 &2M17532599-2053304 & 3896.9$\pm$6 &  0.91$\pm$0.03 &  $-$0.87$\pm$0.01 & 2.10 &c5 $-$0.05$\pm$0.03 & +0.25$\pm$0.03 & +0.30$\pm$0.03   & +0.85$\pm$0.05 & +0.60$\pm$0.05 &  +0.54/{\bf 0.60}$\pm$0.03  &$-$0.20$\pm$0.03    &$-$0.60$\pm$0.05& \\   
c6 &2M18044663-3132174 & 3832.6$\pm$6 &  0.92$\pm$0.03 &  $-$0.90$\pm$0.01 & 2.22 &c6 +0.10$\pm$0.03 & +0.35$\pm$0.05 & +0.35$\pm$0.05     & +0.45$\pm$0.10 & +0.42$\pm$0.05 &  +0.33$\pm$0.07             &  +0.30$\pm$0.03    & +0.28$\pm$0.05& \\    
c7 &2M18080306-3125381 & 4310.0$\pm$12 & 1.57$\pm$0.04 &  $-$0.90$\pm$0.02 & 1.48 &c7 $-$0.07$\pm$0.03 & +0.25$\pm$0.03 & +0.35$\pm$0.03   & +0.50$\pm$0.10 & +0.25$\pm$0.10 &  +0.23$\pm$0.11             &  +0.20$\pm$0.03    & +0.09$\pm$0.07& \\    
c8 &2M18195859-1912513 & 4102.0$\pm$10 & 1.05$\pm$0.04 &  $-$1.24$\pm$0.01 & 1.78 &c8 $-$0.20$\pm$0.03 & +0.40$\pm$0.03 & +0.25$\pm$0.03   & +0.40$\pm$0.10 & +0.50$\pm$0.10 &  +0.79$\pm$0.11             &  +0.25$\pm$0.03    & +0.32$\pm$0.07& \\    
c9 &2M17190320-2857321 & 4139.6$\pm$12 & 1.19$\pm$0.05 &  $-$1.20$\pm$0.01 & 1.83 &c9 $-$0.30$\pm$0.05 & +0.20$\pm$0.08 & +0.50$\pm$0.05   &    ---         & +0.60$\pm$0.05 &  +0.31$\pm$0.16             &  +0.20$\pm$0.10    & +0.21$\pm$0.09& \\    
c10 &2M17224443-2343053& 4058.3$\pm$7 &  1.02$\pm$0.03 &  $-$0.88$\pm$0.01 & 1.97 &c10 $-$0.10$\pm$0.03 & +0.20$\pm$0.03 & +0.37$\pm$0.03  & +0.75$\pm$0.05 & +0.30$\pm$0.05 &  +0.53/{\bf 0.42}$\pm$0.02  &  +0.25$\pm$0.03    & +0.23$\pm$0.05& \\    
c11 &2M17292082-2126433& 3983.4$\pm$7 &  0.78$\pm$0.04 &  $-$1.27$\pm$0.01 & 2.59 &c11  $-$0.20$\pm$0.03 & +1.10$\pm$0.05 & +0.75$\pm$0.05 & +0.40$\pm$0.05 & +0.35$\pm$0.10 &  +0.24$\pm$0.09             &  +0.50$\pm$0.03    & +0.19$\pm$0.06& \\    
c12 &2M17323787-2023013& 3865.7$\pm$6 &  1.03$\pm$0.03 &  $-$0.85$\pm$0.01 & 1.94 &c12 +0.07$\pm$0.03 & +0.20$\pm$0.03 & +0.40$\pm$0.03    & +0.70$\pm$0.05 & +0.70$\pm$0.10 &  +0.70/{\bf 0.49}$\pm$0.018  &  +0.25$\pm$0.03    & +0.24$\pm$0.05& \\    
c13 &2M17330730-2407378& 4042.5$\pm$9 &  0.25$\pm$0.05 &  $-$1.90$\pm$0.01 & 1.88 &c13 $-$0.60$\pm$0.03 & +0.50$\pm$0.03 & +0.30$\pm$0.03  & +0.40$\pm$0.10 & +0.20$\pm$0.10 &$-$0.30$\pm$0.15             &  +0.10$\pm$0.10    & +0.09$\pm$0.08& \\    
c14 &2M18023156-2834451& 3617.4$\pm$6 &  0.42$\pm$0.04 &  $-$1.19$\pm$0.01 & 3.02 &c14 $-$0.05$\pm$0.03 & +0.20$\pm$0.03 & +0.30$\pm$0.03  & +0.97$\pm$0.05 & +0.32$\pm$0.10 &  ---                        &  +0.10$\pm$0.03    & +0.20$\pm$0.06& \\    
c15 &2M17291778-2602468& 3844.3$\pm$7 &  0.71$\pm$0.04 &  $-$0.99$\pm$0.01 & 2.10 &c15 +0.05$\pm$0.03 & +0.30$\pm$0.03 & +0.38$\pm$0.03    & +0.50$\pm$0.10 & +0.70$\pm$0.05 &  +0.71$\pm$0.08             &  +0.15$\pm$0.03    & +0.23$\pm$0.06& \\    
c16 &2M17310874-2956542& 4175.7$\pm$10 & 1.20$\pm$0.05 &  $-$0.93$\pm$0.01 & 2.07 &c16 $-$0.10$\pm$0.03 & +0.20$\pm$0.03 & +0.30$\pm$0.03  & +0.55$\pm$0.10 & +0.35$\pm$0.05 &  +0.57$\pm$0.10             &  +0.17$\pm$0.03    & +0.08$\pm$0.07& \\    
c17 &2M17382504-2424163& 3880.4$\pm$6 &  0.99$\pm$0.04 &  $-$1.05$\pm$0.01 & 1.55 &c17 $-$0.20$\pm$0.03 & +0.30$\pm$0.03 & +0.40$\pm$0.03  & +0.80$\pm$0.05 & +0.25$\pm$0.10 &$-$0.09$\pm$0.08             &  +0.07$\pm$0.03    & +0.18$\pm$0.05& \\    
c18 &2M17511568-3249403& 3921.2$\pm$9 &  0.98$\pm$0.05 &  $-$0.90$\pm$0.02 & 2.04 &c18 +0.00$\pm$0.03 & +0.00$\pm$0.03 & +0.38$\pm$0.03    & +0.75$\pm$0.10 & +0.36$\pm$0.05 &  +0.89$\pm$0.10             &  +0.25$\pm$0.03    & +0.26$\pm$0.07& \\    
c19 &2M17552681-3334272& 4051.0$\pm$10 & 1.08$\pm$0.05 &  $-$0.89$\pm$0.02 & 1.98 &c19 $-$0.10$\pm$0.03 & +0.00$\pm$0.03 & +0.40$\pm$0.03  & +0.85$\pm$0.05 &  ---           &  +0.27$\pm$0.11             &  +0.25$\pm$0.03    & +0.18$\pm$0.07& \\    
c20 &2M18005152-2916576& 4158.9$\pm$12 & 1.04$\pm$0.06 &  $-$1.02$\pm$0.02 & 2.21 &c20 $-$0.35$\pm$0.03 & +0.20$\pm$0.03 & +0.40$\pm$0.03  & +1.00$\pm$0.05 & +0.55$\pm$0.05 &  +0.68$\pm$0.14             &  +0.05$\pm$0.10    & +0.15$\pm$0.08& \\    
c21 &2M18010424-3126158& 3773.1$\pm$6 &  0.68$\pm$0.03 &  $-$0.83$\pm$0.01 & 2.20 &c21 $-$0.05$\pm$0.03 & +0.00$\pm$0.03 & +0.38$\pm$0.03  & +0.55$\pm$0.05 & +0.32$\pm$0.05 &  +0.32/{\bf 0.38}$\pm$0.03  &  +0.15$\pm$0.03    & +0.17$\pm$0.05& \\    
c22 &2M18042687-2928348& 4164.7$\pm$11 & 0.88$\pm$0.05 &  $-$1.21$\pm$0.02 & 2.14 &c22 $-$0.30$\pm$0.03 & +0.40$\pm$0.08 & +0.30$\pm$0.05  &    ---         & +0.55$\pm$0.05 &  +0.55$\pm$0.13             &  +0.07$\pm$0.10    & +0.17$\pm$0.08& \\    
c23 &2M18052388-2953056& 4252.9$\pm$14 & 0.92$\pm$0.06 &  $-$1.57$\pm$0.02 & 1.92 &c23 $-$0.35$\pm$0.08 & +0.40$\pm$0.10 & +0.40$\pm$0.05  &    ---         & +0.25$\pm$0.10 &  +0.50$\pm$0.19             &  +0.22$\pm$0.20    & +0.50$\pm$0.10& \\    
c24 &2M18142265-0904155& 3920.5$\pm$11 & 1.12$\pm$0.05 &  $-$0.85$\pm$0.02 & 2.13 &c24 +0.00$\pm$0.03 & +0.20$\pm$0.03 & +0.33$\pm$0.03    & +0.45$\pm$0.10 & +0.50$\pm$0.10 &  +0.51$\pm$0.13             &  +0.27$\pm$0.03    & +0.30$\pm$0.08& \\    
c25 &2M17293482-2741164& 4143.5$\pm$8 &  1.03$\pm$0.04 &  $-$1.25$\pm$0.01 & 1.85 &c25 $-$0.20$\pm$0.03 & +0.30$\pm$0.03 & +0.40$\pm$0.03  &$<$+0.40        & +0.30$\pm$0.05 &  +0.57/{\bf 0.33}$\pm$0.10  &  +0.22$\pm$0.03    & +0.22$\pm$0.06& \\    
c26 &2M17341796-3905103& 4163.5$\pm$13 & 1.42$\pm$0.05 &  $-$0.89$\pm$0.02 & 1.84 &c26 +0.00$\pm$0.03 & +0.25$\pm$0.03 & +0.40$\pm$0.03    & +0.60$\pm$0.10 & +0.55$\pm$0.10 &  +0.65$\pm$0.13             &  +0.15$\pm$0.15    & +0.09$\pm$0.08& \\    
c27 &2M17342067-3902066& 4380.4$\pm$16 & 1.40$\pm$0.05 &  $-$0.93$\pm$0.02 & 1.99 &c27 +0.15$\pm$0.06 & +0.20$\pm$0.06 & +0.40$\pm$0.06    & +0.70$\pm$0.10 & +0.25$\pm$0.10 &  +0.62$\pm$0.15             &  +0.11$\pm$0.10    &$-$0.02$\pm$0.09& \\   
c28 &2M17503065-2313234& 3821.4$\pm$6 &  0.98$\pm$0.03 &  $-$0.88$\pm$0.01 & 2.10 &c28 +0.10$\pm$0.03 & +0.10$\pm$0.03 & +0.35$\pm$0.05    &    ---         & +0.40$\pm$0.05 &  +0.35$\pm$0.07             &  +0.17$\pm$0.03    & +0.22$\pm$0.05& \\    
c29 &2M18143710-2650147& 4244.5$\pm$11 & 1.30$\pm$0.04 &  $-$0.92$\pm$0.01 & 1.97 &c29 $-$0.10$\pm$0.08 & +0.10$\pm$0.10 & +0.40$\pm$0.08  &    ---         & +0.25$\pm$0.05 &  +0.42$\pm$0.10             &  +0.25$\pm$0.10    & +0.06$\pm$0.07& \\    
c30 &2M18150516-2708486& 3833.4$\pm$9 &  1.00$\pm$0.04 &  $-$0.83$\pm$0.02 & 2.14 &c30 +0.10$\pm$0.03 & +0.00$\pm$0.05 & +0.30$\pm$0.05    &    ---         & +0.40$\pm$0.05 &  +0.56$\pm$0.10             &  +0.06$\pm$0.03    & +0.19$\pm$0.07& \\    
c31 &2M18344461-2415140& 4297.6$\pm$11 & 1.09$\pm$0.05 &  $-$1.42$\pm$0.01 & 1.83 &c31 $-$0.45$\pm$0.10 & +0.40$\pm$0.10 & +0.25$\pm$0.10  &    ---         & +0.20$\pm$0.10 &  +0.28/{\bf 0.12}$\pm$0.22  &  +0.12$\pm$0.08    & +0.23$\pm$0.08e& \\    
\hline
\noalign{\smallskip}
\hline 
\label{results}
\end{tabular}}
\begin{minipage}{13cm}
\vspace{0.1cm}
\small Notes: [S/Fe] values in bold face are from the Value Added Catalogue (VAC) data derived with BAWLAS;
$^{*}$ star b13 could be uncertain, because the \ion{P}{I} line is at the border of the spectrum just before the wavelength gap.
\end{minipage}
\end{table*}

\end{appendix}

\end{document}